\documentclass[a4paper,12pt, leqno]{article}

\usepackage{amsmath}
\usepackage{amssymb, latexsym}
\usepackage{amsfonts}
\usepackage{makeidx}
\usepackage{amssymb}
\usepackage{color}

\date{}

\newtheorem{lem}{Lemma}[section]
\newtheorem{thm}{Theorem}[section]
\newtheorem{prop}{Proposition}[section]
\newtheorem{cor}{Corollary}[section]

\numberwithin{equation}{section}
\newcommand{\dbar}{d\!\!\!{\lower-0.6ex\hbox{$-$}}\!}
\newcommand{\dslash}{d\!\!\!{\lower-0.6ex\hbox{$-$}}}
\newcommand{\e}{\varepsilon}
\newcommand{\h}{\hbar}
\newcommand{\ott}{\lower-0.4ex\hbox{${\scriptscriptstyle{\otimes}}$}}
\newcommand{\btt}{\lower-0.2ex\hbox{${\scriptscriptstyle{\bullet}}$}}
\newcommand{\ctt}{\lower-0.2ex\hbox{${\scriptscriptstyle{\circ}}$}}
\newcommand{\dtt}{\lower-0.2ex\hbox{${\scriptscriptstyle{\diamond}}$}}
\newcommand{\odt}{\lower-0.4ex\hbox{${\scriptscriptstyle{\odot}}$}}

\begin{document}

\title{Deformation Expression for Elements of Algebras (V)\\
--Diagonal matrix calculus and $*$-special functions--}


\author{
     Hideki Omori\thanks{ Department of Mathematics,
             Faculty of Sciences and Technology,
        Tokyo University of Science, 2641, Noda, Chiba, 278-8510, Japan,
         email: omori@ma.noda.sut.ac.jp}
        \\Tokyo University of Science
\and  Yoshiaki Maeda\thanks{Department of Mathematics,
                Faculty of Science and Technology,
                Keio University, 3-14-1, Hiyoshi, Yokohama,223-8522, Japan,
                email: maeda@math.keio.ac.jp}
          \\Keio University
\and  Naoya Miyazaki\thanks{ Department of Mathematics, Faculty of
Economics, Keio University,  4-1-1, Hiyoshi, Yokohama, 223-8521, Japan,
        email: miyazaki@hc.cc.keio.ac.jp}
        \\Keio University
\and  Akira Yoshioka \thanks{ Department of Mathematics,
          Faculty of Science, Tokyo University of Science,
         1-3, Kagurazaka, Tokyo, 102-8601, Japan,
         email: yoshioka@rs.kagu.tus.ac.jp}
           \\Tokyo University of Science
     }

\maketitle

\tableofcontents



\setcounter{equation}{0}

In this note, we are interested in the $*$-version of various special
functions. Noting that many special functions are 
defined by integrals involving the exponential functions, we define 
$*$-special functions by similar integral formula replacing   
exponential functions by ${*}$-exponential functions. 

As it is seen in previous notes, a ${*}$-exponential function of a
nondegenerate quadratic form has remarkable properties which are not 
seen in ordinary exponential functions. 

To treat these it is convenient to use diagonal matrix
  expressions developed in this note.

\section{Summary of fundamental properties of $*$-exponential functions}

The Weyl algebra $(W_2; *)$ is the algebra generated by $u,v$ with 
$u{*}v{-}v{*}u=-i\h$. 
For an arbitrary fixed $2{\times}2$-complex symmetric matrix 
$K$, we set $\Lambda=K{+}J$ where $J$ is the standard skew-symmetric
matrix  
$J{=}\tiny{
\begin{bmatrix}
0&-1\\
1&0
\end{bmatrix}}$. 
Setting $(u,v){=}(u_1,u_2)$, we define a product ${*}_{_{K}}$ on the space
of polynomials  ${\mathbb C}[u_1,u_2]$ by the formula 
\begin{equation}
 \label{eq:KK}
 f*_{_{K}}g=fe^{\frac{i\h}{2}
(\sum\overleftarrow{\partial_{u_i}}
{\Lambda}{}^{ij}\overrightarrow{\partial_{u_j}})}g
=\sum_{k}\frac{(i\h)^k}{k!2^k}
{\Lambda}^{i_1j_1}\!{\cdots}{\Lambda}^{i_kj_k}
\partial_{u_{i_1}}\!{\cdots}\partial_{u_{i_k}}f\,\,
\partial_{u_{j_1}}\!{\cdots}\partial_{u_{j_k}}g.   
\end{equation}
It is known and not hard to prove that 
$({\mathbb C}[u_1,u_2],*_{_{K}})$ is  
isomorphic to $(W_2; *)$. 
If $K$ is fixed, then every element $A{\in}(W_2; *)$ 
is expressed in the form of ordinary polynomial, which we 
denote by ${:}A{:}_{_K}\in {\mathbb C}[u_1,u_2]$.
Thus, for every $K$, $({\mathbb C}[u_1,u_2],*_{_{K}})$
gives a concrete representation of the Weyl algebra, and the 
product formula \eqref{eq:KK} offers several ways to calculate 
transcendental elements.  

\medskip
All $({\mathbb C}[u_1,u_2],*_{_{K}})$ are mutually isomorphic, that is, there 
is an isomorphism 
$$
I_{_K}^{^{K'}}: ({\mathbb C}[u_1,u_2],*_{_{K}})\to ({\mathbb C}[u_1,u_2],*_{_{K'}})
$$
called {\bf intertwiner} for every $K$ and $K'$. This is given by 
$$
I_{_K}^{^{K'}}(f(u_1,u_2))=\Big(\exp\frac{i\h}{4}\sum
(K'_{ij}{-}K_{ij})
\partial_{i}\partial_{j}\Big)f(u_1,u_2).
$$
The intertwiner $I_{_K}^{^{K'}}$ extends for several class of
transcendental elements. This works pretty well for exponential functions of
linear functions, and it is still amenable for exponential functions of
quadratic functions (cf.\cite{OMMY5}, \cite{ommy6}), but in general the extended intertwiner $I_{_K}^{^{K'}}$
is neither continuous nor $1$-to-$1$.  

\medskip
For an element $H_*{\in}(W_2; *)$, the $*$-exponential function of $H_*$
is defined by the evolution equation  
\begin{equation}\label{evol111}
\frac{d}{dt}f_t={:}H_*{:}_{_K}{*}f_t,\quad f_0=1.
\end{equation}
If the real analytic solution exists then the solution is denoted by 
${:}e_*^{tH_*}{:}_{_K}$. If the real analytic solution of the initial
data $g$ 
exists then the solution is denoted by ${:}e_*^{tH_*}{*}g{:}_{_K}$. 

In what follows we mainly concern the case
$H_*=\frac{1}{i\h}u{\ctt}v$, where $u{\ctt}v{=}\frac{1}{2}(u{*}v{+}v{*}u)$. 

For  
$K=\left[\begin{array}{cc}\delta & c \\ c & \delta'\end{array}\right]$, 
by setting 
$\Delta{=}e^t{+}e^{-t}{-}c(e^t{-}e^{-t})$,  
the solution of \eqref{evol111} is given by   
\begin{equation}\label{genericparam00}
{:}e_*^{t\frac{1}{i\h}2u{\ctt}v}{:}_{_{K}}{=}
\frac{2}{\sqrt{\Delta^2{-}(e^t{-}e^{-t})^2\delta\delta'}} 
\,\,e^{\frac{1}{i\h}
\frac{e^t-e^{-t}}{\Delta^2{-}(e^t{-}e^{-t})^2\delta\delta'}
\big((e^t-e^{-t})(\delta' u^2{+}\delta v^2){+}2\Delta uv\big)}.
\end{equation}

As ${:}e_*^{z\frac{1}{i\h}2u{\ctt}v}{:}_{_K}$ has double
branched singularities in general, we have to prepare two $\pm$ sheets with slits. 
Hence, we have two origins $0_+$, $0_-$. 
First, note that 
${:}e_*^{0\frac{1}{i\h}2u{\ctt}v}{:}_{_{K}}{=}\sqrt{1}$.
Thus, we set 
\begin{equation}\label{seteval}
{:}e_*^{0_+\frac{1}{i\h}u{\ctt}v}{:}_{_K}{=}1,\quad 
{:}e_*^{0_-\frac{1}{i\h}u{\ctt}v}{:}_{_K}{=}-1.
\end{equation}
Note that 
${:}e_*^{\pm\pi i\frac{1}{i\h}2u{\ctt}v}{:}_{_{K}}{=}\sqrt{1}$, 
but this is {\it not} an absolute scalar.
The $\pm$ sign depends on $K$ and the path form $0$ to $\pi i$ by setting 
${:}e_*^{0\frac{1}{i\h}2u{\ctt}v}{:}_{_{K}}{=}1$. 
As this is a scalar-like element belonging to a one parameter subgroup, 
we call it a $q$-{\it scalar}. 

If $t=\pm\frac{\pi i}{2}$ in \eqref{genericparam00}, then 
${:}e_*^{\pm\frac{\pi i}{2}\frac{1}{i\h}2u{\ctt}v}{:}_{_{K}}$ 
is called the {\bf polar element} and denoted by 
${\e}_{00}$ \cite{OMMY4}, \cite{OMMY5}: 
\begin{equation}\label{polar}
{:}{\e}_{00}{:}_{_K}=\frac{1}{\sqrt{c^2{-}\delta\delta'}}
e^{\frac{1}{i\h}\frac{1}{c^2{-}\delta\delta'}
(\delta' u^2{+}\delta v^2 {-}2cuv)}.
\end{equation}

\medskip
Let $H\!ol({\mathbb C}^2)$ be the space of all
holomorphic functions of $(u,v)\in {\mathbb C}^2$ with the uniform
convergent topology on each compact subset. 
In \cite{ommy6}, we summarize properties for generic expression parameters $K$ as follows:

\noindent
({\bf a})\,\,
$:e_*^{z\frac{2}{i\h}u{\ctt}v}$ has no singular point on the real axis and the 
pure imaginary axis.

\noindent
({\bf b})\,\,$e_*^{z\frac{2}{i\h}u{\ctt}v}$ is rapidly decreasing along any  
line parallel to the real line.

\noindent
({\bf c})\,\,$e_*^{z\frac{2}{i\h}u{\ctt}v}$ is a $H\!ol({\mathbb  C}^2)$-valued 
$2\pi i$-periodic function, i.e. 
${:}e_*^{(z{+}2\pi i)\frac{2}{i\h}u{\ctt}v}{:}_{_K}={:}e_*^{z\frac{2}{i\h}u{\ctt}v}{:}_{_K}$. 
More precisely, 

it is $\pi i$-periodic or alternating $\pi i$-periodic depending on the real part of $z$ 

\noindent
({\bf d})\,\,As a result, $e_*^{z\frac{2}{i\h}u{\ctt}v}$ must have periodic singular
points. But the singular points are double 

branched. Hence $e_*^{z\frac{2}{i\h}u{\ctt}v}$ is double valued with the sign
ambiguity. Singular point set $\Sigma_K$ is 

distributed $\pi i$-periodically along the two lines parallel to the imaginary
axis.

\noindent
({\bf e}) By requesting $1$ at $z=0$,
i.e. ${:}e_*^{0\frac{2}{i\h}u{\ctt}v}{:}_{_K}=1$, the value 
${:}e_*^{[0{\sim}z]\frac{2}{i\h}u{\ctt}v}{:}_{_K}$ is determined
uniquely, where 
$[0{\sim}z]$ is a path from $0$ to $z$ avoiding $\Sigma_K$ 
and evaluating at $z$. Thus in spite of the double 
valued nature, the exponential law 
$$
e_*^{z\frac{2}{i\h}u{\ctt}v}{*}e_*^{z'\frac{2}{i\h}u{\ctt}v}=
e_*^{(z{+}z')\frac{2}{i\h}u{\ctt}v}
$$

holds under the calculation such that $\sqrt{a}\sqrt{b}=\sqrt{ab}$.

\subsection{Exchanging interval and idempotent elements} 
As the pattern of periodicity depends on how the 
circle $\{re^{i\theta}; \theta\in \mathbb R\}$ 
round the singular points, it depends delicately on $K$. 
As it is mentioned in ({\bf d}), there is two lines
${\rm{Re}\,z}{=}a$, ${\rm{Re}\,z}{=}b$ on which singular points 
of $e_*^{z\frac{2}{i\h}u{\ctt}v}$ are sitting $\pi i$-periodically.
The interval $(a,b)$ depends on the expression parameter $K$. We
denote this by $I_{\ctt}(K)=(a,b)$, and we call it  
the (sheet) {\bf exchanging interval} of
$e_*^{z\frac{1}{i\h}2u{\ctt}v}$. 

\bigskip
As the exponential law shows that 
$e_*^{z\frac{1}{i\h}2u{*}v}$, $e_*^{z\frac{1}{i\h}2v{*}u}$
have singular points on the same two lines, 
$I_{\ctt}(K)$ is called also the exchanging interval of these. 

The periodicity of ${:}e_*^{(s{+}it)\frac{1}{i\h}2u{\ctt}v}{:}_{_K}$
w.r.t. $t$ depends on $s$ as follows:

If $s{<}a$, or $s>b$, then $e_*^{(s{+}it)\frac{1}{i\h}2u{\ctt}v}$ is
alternating $\pi$-periodic, and if $a{<}s{<}b$, then 
$e_*^{(s{+}it)\frac{1}{i\h}2u{\ctt}v}$ is $\pi$-periodic.   

\medskip
  
There are three disjoint open subsets 
${\mathfrak K}_{\pm}$ and ${\mathfrak K}_0$ of the space of
$2{\times}2$ 
symmetric of all expression parameters such that 

\bigskip
\noindent
(i) ${\mathfrak K}_{+}\cup{\mathfrak K}_{-}\cup{\mathfrak K}_{0}$ is
dense. 

\medskip
\noindent
(ii) If $K\in {\mathfrak K}_{0}$, then $a{<}0{<}b$. 
${:}{\e}_{00}^2{:}_{_K}=1$, and 
$\varpi_*(0)=\frac{1}{4\pi}\int_{-2\pi}^{2\pi}e_*^{it\frac{1}{i\h}u{\ctt}v}dt$
is called the {\bf pseudo-vacuum}. 
Cauchy's integral theorem shows that 
$\frac{1}{4\pi}\int_{-2\pi}^{2\pi}e_*^{(s{+}it)\frac{1}{i\h}u{\ctt}v}dt$
is independent of $s$ whenever $a<s<b$. 

\medskip
\noindent
(iii) If $K\in {\mathfrak K}_{+}$, then $0{<}a$ and 
${:}{\e}_{00}^2{:}_{_K}=-1$.
$\varpi_{00}=\frac{1}{4\pi}\int_{-2\pi}^{2\pi}e_*^{it\frac{1}{i\h}(u{\ctt}v{-}\frac{1}{2})}dt$
is called the {\bf vacuum}. 
Cauchy's integral theorem shows that 
$\frac{1}{4\pi}\int_{-2\pi}^{2\pi}e_*^{(s{+}it)(\frac{1}{i\h}u{\ctt}v{-}\frac{1}{2})}dt$
is independent of $s$ whenever $s<a$.

\medskip
\noindent
(iv) If $K\in {\mathfrak K}_{-}$, then $b{<}0$ and 
${:}{\e}_{00}^2{:}_{_K}=-1$. 
$\overline{\varpi}_{00}
=\frac{1}{4\pi}\int_{-2\pi}^{2\pi}e_*^{it\frac{1}{i\h}(u{\ctt}v{+}\frac{1}{2})}dt$
is called the {\bf bar-vacuum}. 
Cauchy's integral theorem shows that 
$\frac{1}{4\pi}\int_{-2\pi}^{2\pi}e_*^{(s{+}it)(\frac{1}{i\h}u{\ctt}v{+}\frac{1}{2})}dt$
is independent of $s$ whenever $b<s$. 

\bigskip
\noindent
{\bf Idempotent elements}\,\,

It is not hard to show the idempotent properties of 
$\varpi_{00}, \,\, \overline{\varpi}_{00}, \,\, \varpi_*(0)$.
As the double integral 
$\frac{1}{4\pi}\int_{-2\pi}^{2\pi}e_*^{(it{+}i\tau)\frac{1}{i\h}(u{\ctt}v{-}\frac{1}{2})}dtd\tau$
is welldefined, the exponential law and the change of variables gives
that if $0{<}a$, then   
$$
\frac{1}{4\pi}\int_{-2\pi}^{2\pi}e_*^{it\frac{1}{i\h}(u{\ctt}v{-}\frac{1}{2})}dt{*}
\frac{1}{4\pi}\int_{-2\pi}^{2\pi}e_*^{i\tau\frac{1}{i\h}(u{\ctt}v{-}\frac{1}{2})}d\tau{=}
\frac{1}{4\pi}\int_{-2\pi}^{2\pi}e_*^{it\frac{1}{i\h}(u{\ctt}v{-}\frac{1}{2})}dt.
$$
Others are obtained similarly. These are given by periodic integrals along the imaginary
axis.

\bigskip
Besides these, if $|{\rm{Re}}\,z|<1/2$ then the integral in generic ordered expression 
$$
\int_{-\infty}^{\infty}{:}e^{t(z{+}\frac{1}{i\h}2u{\ctt}v{+}i\xi)}{:}_{_K}ds{=}
\int_{-\infty}^{\infty}\frac{e^{-it(iz{-}\xi)}}
{\sqrt{\Delta^2{-}(e^t{-}e^{-t})^2\delta\delta'}} 
\,\,e^{\frac{1}{i\h}
\frac{e^t-e^{-t}}{\Delta^2{-}(e^t{-}e^{-t})^2\delta\delta'}
\big((e^t-e^{-t})(\delta' u^2{+}\delta v^2){+}2\Delta uv\big)}dt.
$$ 
is rapidly decreasing w.r.t. $\xi\in{\mathbb R}$. We denote this by 
\begin{equation}\label{deltaintg}
\int_{-\infty}^{\infty}e^{s(z{+}\frac{1}{i\h}u{\ctt}v+i\xi)}ds=
\int_{-\infty}^{\infty}e^{-is(iz{+}\frac{1}{\h}u{\ctt}v{-}\xi)}ds=
\delta_*(iz{+}\frac{1}{\h}u{\ctt}v{-}\xi),\quad |{\rm{Re}}\,z|<1/2.
\end{equation}
Moreover, the property similar to idempotency  
\begin{equation}\label{deltaidem}
\delta_*(iz{+}\frac{1}{\h}u{\ctt}v{-}\xi)
{*}\delta_*(iz{+}\frac{1}{\h}u{\ctt}v{-}\xi')=
\delta(\xi-\xi')\delta_*(iz{+}\frac{1}{\h}u{\ctt}v{-}\xi),\quad |{\rm{Re}}\,z|<1/2,
\end{equation}
is proved directly as follows: As the double integral is welldefined, we have
$$
\begin{aligned}
\delta_*&(iz{+}\frac{1}{\h}u{\ctt}v{-}\xi){*}\delta_*(iz{+}\frac{1}{\h}u{\ctt}v{-}\xi')=
\iint 
e_*^{-it(iz{+}\frac{1}{\h}u{\ctt}v{-}\xi)}{*}e_*^{-is(iz{+}\frac{1}{\h}u{\ctt}v{-}\xi')}dtds\\
=&
\iint e^{it\xi{+}is\xi'}e_*^{-i(t{+}s)(iz{+}\frac{1}{\h}u{\ctt}v)}dtds=
\iint e^{is(\xi'{-}\xi)}e_*^{-i\sigma(iz{-}\xi{+}\frac{1}{\h}u{\ctt}v)}dsd\sigma
=\delta(\xi'{-}\xi)\delta_*(iz{+}\frac{1}{\h}u{\ctt}v{-}\xi).
\end{aligned}
$$

Similar to the properties (ii), (iii) and (iv), 
$\int_{-\infty}^{\infty}e^{(s{+}it)(z{+}\frac{1}{i\h}u{\ctt}v+i\xi)}ds$
is independent of $t$ whenever $|t|$ is very small by 
Cauchy's integral theorem. Thus, by
differentiating these we have 
$$
(\frac{1}{i\h}u{\ctt}v){*}\delta_*(\frac{1}{\h}u{\ctt}v)=0,\quad 
(\frac{1}{i\h}u{\ctt}v{-}\frac{1}{2}){*}\varpi_{00}=0,\quad 
(\frac{1}{i\h}u{\ctt}v{+}\frac{1}{2}){*}\overline{\varpi}_{00}=0,\quad 
(\frac{1}{i\h}u{\ctt}v){*}{\varpi}_{*}(0)=0.
$$
But note that differentiations are not by complex variable,
but by the real/pure imaginary variables.  Now, recall the equality
given in \cite{ommy6}: In generic ordered expression $K$, let
$I_{\ctt}(K)=(a,b)$ be the exchanging interval. We have then
\begin{equation}\label{1steqeq}
\begin{aligned}
&\frac{1}{4\pi}\int_{-2\pi}^{2\pi}{:}e_*^{(s{+}i\sigma)\frac{1}{i\h}u{*}v}{:}_{_K}d\sigma
=\left\{
\begin{matrix}
{:}\varpi_{00}{:}_{_K},&  s<a\\
0,& a<s
\end{matrix}
\right.\\
&\frac{1}{4\pi}\int_{-2\pi}^{2\pi}{:}e_*^{(s{+}i\sigma)\frac{1}{i\h}v{*}u}{:}_{_K}d\sigma
=\left\{
\begin{matrix}
0,&  s<b\\
{:}\overline{\varpi}_{00}{:}_{_K},& b<s.
\end{matrix}
\right.
\end{aligned}
\end{equation}
Hence we see 
\begin{equation}\label{vectdist}
\begin{aligned}
{:}(\frac{1}{i\h}u{\ctt}v{-}\frac{1}{2})&{*}
\frac{1}{4\pi}\int_{-2\pi}^{2\pi}{:}e_*^{(s{+}i\sigma)(\frac{1}{i\h}u{\ctt}v{-}\frac{1}{2})}{:}_{_K}d\sigma
{=}{-}\delta(s{-}a){:}\varpi_{00}{:}_{_K}\\ 
{:}(\frac{1}{i\h}u{\ctt}v{+}\frac{1}{2})&{*}
\frac{1}{4\pi}\int_{-2\pi}^{2\pi}{:}e_*^{(s{+}i\sigma)(\frac{1}{i\h}u{\ctt}v{+}\frac{1}{2})}{:}_{_K}d\sigma
{=}\delta(s{-}b){:}\overline{\varpi}_{00}{:}_{_K}
\end{aligned}
\end{equation}
It is easy to see that first equality of \eqref{1steqeq} is the solution of 
$$
\frac{d}{ds}f(s)={-}\delta(s{-}a)\varpi_{00}{*}f(s),\quad f(-\infty)=\varpi_{00}.
$$
Similarly, we have  
\begin{equation*}
\frac{1}{4\pi}\int_0^{4\pi}{:}e_*^{(s{+}it)(\frac{1}{i\h}u{\ctt}v)}{:}_{_K}dt=
\left\{
\begin{matrix}
0,& s<a&{ }\\
{:}{\tilde D}_0{:}_{_K},&a<s<b,\\
0,& b<s&{ }
\end{matrix}
\right.
\end{equation*}
where 
${\tilde D}_0=\frac{1}{4\pi}\int_{-2\pi}^{2\pi}e_{*}^{(s{+}it)\frac{1}{i\h}u{\ctt}v}$ 
which is independent of $s$, $a<s<b$ by Cauchy's integration theorem. 
Furthermore, ${\tilde D}_0=\varpi_*(0)$ if  $K\in{\mathfrak K}_0$.
This gives 
\begin{equation}\label{vectdist2}
{:}(\frac{1}{i\h}u{\ctt}v){*}\frac{1}{4\pi}\int_0^{4\pi}{:}e_*^{(s{+}it)(\frac{1}{i\h}u{\ctt}v)}dt{:}_{_K}
{=}(\delta(s{-}a){-}\delta(s{-}b)){:}{\tilde D}_0{:}_{_K}
\end{equation}

\medskip
To consider the discontinuity of ${:}\int_{\mathbb R}e_*^{(s{+}it)(\frac{1}{i\h}u{\ctt}v)}ds{:}_{_K}$
with respect to $t$, we have to fix several notations about singular
points as in the r.h.s. figure:
 
\vspace{.5cm}
\noindent
\unitlength 0.1in
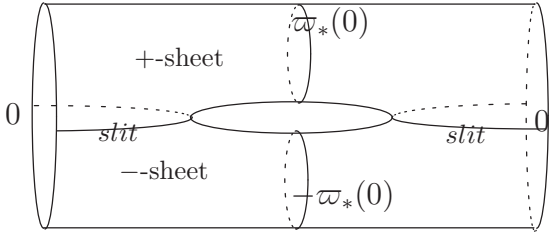
\begin{picture}( 28.1100, 12.7100)(  8.1000,-21.0100)
%
\special{pn 8}%
\special{ar 3562 1498 60 588  4.7123890 6.2831853}%
\special{ar 3562 1498 60 588  0.0000000 1.5707963}%
%
\special{pn 8}%
\special{ar 1016 1502 60 588  4.7123890 6.2831853}%
\special{ar 1016 1502 60 588  0.0000000 1.5707963}%
%
\special{pn 8}%
\special{ar 1010 1498 60 588  1.5707963 4.7123890}%
%
\special{pn 8}%
\special{pa 1016 916}%
\special{pa 3554 916}%
\special{fp}%
\special{pa 3548 2084}%
\special{pa 1010 2084}%
\special{fp}%
%
\special{pn 8}%
\special{ar 3554 1508 46 594  1.4442042 1.4817629}%
\special{ar 3554 1508 46 594  1.5944389 1.6319976}%
\special{ar 3554 1508 46 594  1.7446737 1.7822324}%
\special{ar 3554 1508 46 594  1.8949084 1.9324671}%
\special{ar 3554 1508 46 594  2.0451432 2.0827019}%
\special{ar 3554 1508 46 594  2.1953779 2.2329366}%
\special{ar 3554 1508 46 594  2.3456127 2.3831713}%
\special{ar 3554 1508 46 594  2.4958474 2.5334061}%
\special{ar 3554 1508 46 594  2.6460821 2.6836408}%
\special{ar 3554 1508 46 594  2.7963169 2.8338756}%
\special{ar 3554 1508 46 594  2.9465516 2.9841103}%
\special{ar 3554 1508 46 594  3.0967864 3.1343450}%
\special{ar 3554 1508 46 594  3.2470211 3.2845798}%
\special{ar 3554 1508 46 594  3.3972558 3.4348145}%
\special{ar 3554 1508 46 594  3.5474906 3.5850493}%
\special{ar 3554 1508 46 594  3.6977253 3.7352840}%
\special{ar 3554 1508 46 594  3.8479601 3.8855188}%
\special{ar 3554 1508 46 594  3.9981948 4.0357535}%
\special{ar 3554 1508 46 594  4.1484296 4.1859882}%
\special{ar 3554 1508 46 594  4.2986643 4.3362230}%
\special{ar 3554 1508 46 594  4.4488990 4.4864577}%
\special{ar 3554 1508 46 594  4.5991338 4.6366925}%
%
\special{pn 8}%
\special{ar 2292 1508 522 82  3.9327711 6.2831853}%
\special{ar 2292 1508 522 82  0.0000000 3.9269908}%
%
\special{pn 8}%
\special{ar 2332 1176 60 256  4.7123890 6.2831853}%
\special{ar 2332 1176 60 256  0.0000000 1.5707963}%
%
\special{pn 8}%
\special{ar 2318 1834 60 256  4.7123890 6.2831853}%
\special{ar 2318 1834 60 256  0.0000000 1.5707963}%
%
\special{pn 8}%
\special{ar 2332 1176 60 256  1.5707963 1.6469868}%
\special{ar 2332 1176 60 256  1.8755582 1.9517487}%
\special{ar 2332 1176 60 256  2.1803201 2.2565106}%
\special{ar 2332 1176 60 256  2.4850820 2.5612725}%
\special{ar 2332 1176 60 256  2.7898439 2.8660344}%
\special{ar 2332 1176 60 256  3.0946059 3.1707963}%
\special{ar 2332 1176 60 256  3.3993678 3.4755582}%
\special{ar 2332 1176 60 256  3.7041297 3.7803201}%
\special{ar 2332 1176 60 256  4.0088916 4.0850820}%
\special{ar 2332 1176 60 256  4.3136535 4.3898439}%
\special{ar 2332 1176 60 256  4.6184154 4.6946059}%
%
\special{pn 8}%
\special{ar 2318 1834 60 256  1.5707963 1.6467457}%
\special{ar 2318 1834 60 256  1.8745938 1.9505432}%
\special{ar 2318 1834 60 256  2.1783913 2.2543406}%
\special{ar 2318 1834 60 256  2.4821887 2.5581381}%
\special{ar 2318 1834 60 256  2.7859862 2.8619356}%
\special{ar 2318 1834 60 256  3.0897837 3.1657330}%
\special{ar 2318 1834 60 256  3.3935811 3.4695305}%
\special{ar 2318 1834 60 256  3.6973786 3.7733280}%
\special{ar 2318 1834 60 256  4.0011761 4.0771254}%
\special{ar 2318 1834 60 256  4.3049735 4.3809229}%
\special{ar 2318 1834 60 256  4.6087710 4.6847204}%
%
\special{pn 8}%
\special{ar 1010 1508 768 70  6.2295782 6.2831853}%
\special{ar 1010 1508 768 70  0.0000000 1.4833099}%
%
\special{pn 8}%
\special{ar 3580 1508 766 72  3.0976827 3.1263566}%
\special{ar 3580 1508 766 72  3.2123781 3.2410519}%
\special{ar 3580 1508 766 72  3.3270734 3.3557473}%
\special{ar 3580 1508 766 72  3.4417688 3.4704426}%
\special{ar 3580 1508 766 72  3.5564641 3.5851379}%
\special{ar 3580 1508 766 72  3.6711594 3.6998333}%
\special{ar 3580 1508 766 72  3.7858548 3.8145286}%
\special{ar 3580 1508 766 72  3.9005501 3.9292240}%
\special{ar 3580 1508 766 72  4.0152455 4.0439193}%
\special{ar 3580 1508 766 72  4.1299408 4.1586146}%
\special{ar 3580 1508 766 72  4.2446361 4.2733100}%
\special{ar 3580 1508 766 72  4.3593315 4.3880053}%
\special{ar 3580 1508 766 72  4.4740268 4.5027007}%
\special{ar 3580 1508 766 72  4.5887222 4.6173960}%
%
\special{pn 8}%
\special{ar 3562 1508 748 60  1.5071264 3.0603763}%
%
\special{pn 8}%
\special{ar 1010 1508 764 62  4.6482256 4.6773519}%
\special{ar 1010 1508 764 62  4.7647305 4.7938567}%
\special{ar 1010 1508 764 62  4.8812353 4.9103616}%
\special{ar 1010 1508 764 62  4.9977402 5.0268664}%
\special{ar 1010 1508 764 62  5.1142451 5.1433713}%
\special{ar 1010 1508 764 62  5.2307499 5.2598761}%
\special{ar 1010 1508 764 62  5.3472548 5.3763810}%
\special{ar 1010 1508 764 62  5.4637596 5.4928858}%
\special{ar 1010 1508 764 62  5.5802645 5.6093907}%
\special{ar 1010 1508 764 62  5.6967693 5.7258955}%
\special{ar 1010 1508 764 62  5.8132742 5.8424004}%
\special{ar 1010 1508 764 62  5.9297790 5.9589053}%
\special{ar 1010 1508 764 62  6.0462839 6.0754101}%
\special{ar 1010 1508 764 62  6.1627887 6.1885323}%
\put(27.0000,-9.3000){\makebox(0,0)[rt]{$\varpi_{*}(0)$}}%
\put(22.9000,-20.0000){\makebox(0,0)[lb]{$-\varpi_{*}(0)$}}%
\put(8.1000,-15.4000){\makebox(0,0)[lb]{$0$}}%
\put(35.5000,-15.8000){\makebox(0,0)[lb]{$0$}}%
\put(12.9000,-16.3000){\makebox(0,0)[lb]{\footnotesize{$slit$}}}%
\put(30.9000,-16.4000){\makebox(0,0)[lb]{\footnotesize{$slit$}}}%
\put(14.8000,-12.5000){\makebox(0,0)[lb]{\footnotesize{$+$-sheet}}}%
\put(14.0000,-18.4000){\makebox(0,0)[lb]{\footnotesize{$-$-sheet}}}%
\end{picture}%
\hfill
\unitlength 0.1in
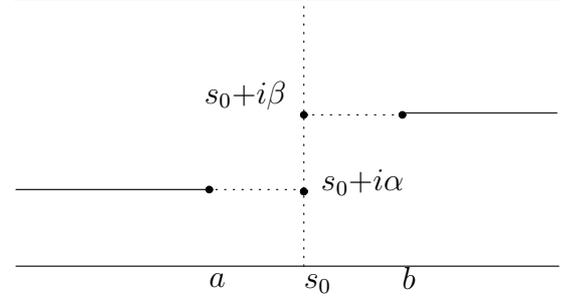
\begin{picture}( 28.2000, 14.0000)(  4.0000,-17.9000)
%
\special{pn 8}%
\special{pa 400 390}%
\special{pa 3220 390}%
\special{fp}%
\special{pa 3220 1790}%
\special{pa 410 1790}%
\special{fp}%
\special{pa 410 1390}%
\special{pa 1410 1390}%
\special{fp}%
\special{pa 2410 990}%
\special{pa 3210 990}%
\special{fp}%
%
\special{pn 8}%
\special{pa 1900 390}%
\special{pa 1900 1790}%
\special{dt 0.045}%
\special{pa 2410 1000}%
\special{pa 1900 1000}%
\special{dt 0.045}%
\special{pa 1900 1390}%
\special{pa 1400 1390}%
\special{dt 0.045}%
%
\special{pn 20}%
\special{sh 1}%
\special{ar 1410 1390 10 10 0  6.28318530717959E+0000}%
\special{sh 1}%
\special{ar 1900 1000 10 10 0  6.28318530717959E+0000}%
\special{sh 1}%
\special{ar 2410 1000 10 10 0  6.28318530717959E+0000}%
\special{sh 1}%
\special{ar 1900 1400 10 10 0  6.28318530717959E+0000}%
\special{sh 1}%
\special{ar 1900 1400 10 10 0  6.28318530717959E+0000}%
\put(19.0000,-19.3000){\makebox(0,0)[lb]{$s_0$}}%
\put(19.9000,-14.2000){\makebox(0,0)[lb]{$s_0{+}i\alpha$}}%
\put(18.1000,-9.7000){\makebox(0,0)[rb]{$s_0{+}i\beta$}}%
\put(14.1000,-19.0000){\makebox(0,0)[lb]{$a$}}%
\put(24.1000,-19.1000){\makebox(0,0)[lb]{$b$}}%
\end{picture}%

It is not hard to see 
\begin{equation}\label{vectdist4}
{:}(\frac{1}{i\h}u{\ctt}v){*}\int_{\mathbb R}e_*^{(s{+}it)(\frac{1}{i\h}u{\ctt}v)}ds{:}_{_K}=
(\delta(t{-}\alpha)-\delta(t{-}\beta))\varpi_*(0)
\end{equation}

\subsection{Matrix elements}
Three idempotent elements $\varpi_{00}$, ${\overline{\varpi}}_{00}$
and $\varpi_{*}(0)$ give matrix elements respectively. 

\begin{prop}
  \label{vacvac4}
In generic ordered expressions,
$E_{p,q}=
\frac{1}{\sqrt{p!q!(i\h)^{p{+}q}}}u^p{*}\varpi_{00}{*}v^q$ 
is the 
$(p,q)$-matrix element, that is 
$E_{p,q}{*}E_{r,s}=\delta_{q,r}E_{p,s}$. The $K$-expression 
${:}E_{p,q}{:}_{_K}$ of $E_{p,q}$ will be denoted by 
$E_{p,q}(K)$. Note that $E_{0,0}(K){=}{:}\varpi_{00}{:}_{_K}$.
\end{prop}

\begin{prop}
\label{barvacvac4}
${\overline E}_{p,q}=
\frac{\sqrt{-1}^{p+q}}{\sqrt{p!q!(i\h)^{p{+}q}}}
v^p{*}{\overline{\varpi}}_{00}{*}u^q$ is the 
$(p,q)$-matrix element. 
 The $K$-expression of 
$\overline{E}_{p,q}$ will be denoted by $\overline{E}_{p,q}(K)$.
Note that $\overline{E}_{0,0}(K){=}{:}{\overline{\varpi}}_{00}{:}_{_K}$.
\end{prop}

\bigskip
Different from the ordinary 
vacuum or bar-vacuum, 
we see $u{*}\varpi_*(0){\not=}0$, 
$v{*}\varpi_*(0){\not=}0$. But note 
that the bumping identity gives 
\begin{equation}\label{slide}
u^n{*}e_*^{it(\frac{1}{i\h}u{\ctt}v)}{=}
e_*^{it(\frac{1}{i\h}u{\ctt}v{-}n)}{*}u^n,
 \quad 
v^n{*}e_*^{it(\frac{1}{i\h}u{\ctt}v)}{=}
e_*^{it(\frac{1}{i\h}u{\ctt}v{+}n)}{*}v^n.
\end{equation}
It is convenient to use the convention    
\begin{equation}\label{pmconvention}
\zeta^k=
\left\{
\begin{matrix}
u^k,& k\geq 0\\
v^{|k|},& k<0,
\end{matrix}
\right.\qquad 
{\hat\zeta}^{\ell}=
\left\{
\begin{matrix}
v^\ell,& \ell\geq 0\\
u^{|\ell|},& \ell<0,
\end{matrix}
\right.
\end{equation}

\begin{prop}\label{quasivac}
If $K{\in}{\mathfrak K}_0$, then    
${:}e_*^{it(\frac{1}{i\h}u{\ctt}v)}{:}_{_K}$ is
$2\pi$-periodic and 
$$
D_{k,\ell}(K)= 
\frac{1}
{\sqrt{(\frac{1}{2})_{k}(\frac{1}{2})_{\ell}(i\h)^{|k|+|\ell|}}}
{\zeta}^k{*}{:}\varpi_*(0){:}_{_K}{*}{\hat\zeta}^{\ell}, \quad 
{:}\varpi_*(0){:}_{_K}{=}
\frac{1}{2\pi}\!\!\int_0^{2\pi}\!\!\!
{:}e_*^{it(\frac{1}{i\h}u{\ctt}v)}dt{:}_{_K}dt
$$
are matrix elements for every $k,\ell{\in}{\mathbb Z}$,  
where $(a)_n=a(a{+}1)\cdots(a{+}n{-1})$, $(a)_0=1$, and extending convention
 $(a)_{-n}=(a{-}1)(a{-}2)\cdots(a{-}n)$. 
 Note that 
$D_{n,n}(K){=}\frac{1}{2\pi}\int_0^{2\pi}{:}e_*^{it(\frac{1}{i\h}u{\ctt}v{-}n)}{:}_{_K}dt$.
\end{prop}

\noindent
{\bf Proof}\,\, 
As the exponential law shows that the double integral 
$\iint(e_*^{it(\frac{1}{i\h}u{\ctt}v)}{*}e_*^{i\tau(\frac{1}{i\h}u{\ctt}v)})dtd\tau$ 
is welldefined, we see that 
if $k{\not=}\ell$, then the change of variables gives 
$$
\int_{0}^{2\pi}
e_*^{is(\frac{1}{i\h}u{\ctt}v{+}k)}ds
*\!\int_{0}^{2\pi}
e_*^{it(\frac{1}{i\h}u{\ctt}v{+}\ell)}dt
{=}
\int_{0}^{2\pi}e^{it(k{-}\ell)}dt
\!\int_{0}^{2\pi}
e_*^{is(\frac{1}{i\h}u{\ctt}v{+}\ell)}ds{=}0, 
$$ 
and 
$$
\frac{1}{2\pi}
\int_{0}^{2\pi}
e_*^{it(\frac{1}{i\h}u{\ctt}v{+}k)}dt
{*}
\frac{1}{2\pi}
\int_{0}^{2\pi}
e_*^{is(\frac{1}{i\h}u{\ctt}v{+}k)}ds
{=}\frac{1}{2\pi}\int_{0}^{2\pi}
e_*^{it(\frac{1}{i\h}u{\ctt}v{+}k)}dt.
$$ 
It is easy to see that  the $*$-product 
$P(u,v){*}\varpi_*(0){*}Q(u,v)$ by any polynomials 
$P(u,v)$, $Q(u,v)$ is reduced to the shape 
$\phi{*}\varpi_*(0){*}\psi$ where $\phi$, $\psi$ are 
polynomials of single variable $u$ or $v$.
Using \eqref{slide}, we have the desired result. \hfill ${\Box}$

Every element of the Weyl algebra is represented by a matrix
by the next theorems:
\begin{thm}\label{basicthm}
In the space $H\!ol({\mathbb C}^2)$, 
$$\sum_{k=0}^{\infty}E_{k,k}(K){=}1, \,\,\text{if }K{\in}{\mathfrak K}_+, \,\,\,\, 
\sum_{n=-\infty}^{\infty}\!D_{n,n}(K){=}1,\,\text{ if }K{\in}{\mathfrak K}_0,\,\,\,\, 
\sum_{k=0}^{\infty}\overline{E}_{k,k}(K){=}1,\,\,\text{ if } K{\in}{\mathfrak K}_-.
$$
\end{thm}
Precisely, they should be written as
$\sum_{k=0}^{\infty}E_{k,k}(K){=}{:}1{:}_{_K}$ etc. for $1$ here is
not an absolute scalar.

\bigskip
\noindent
{\bf Note}\,\,
We easily see that    
\begin{equation}\label{Ebar E}
E_{p,q}{*}\overline{E}_{r,s}=0=\overline{E}_{r,s}{*}E_{p,q}
\end{equation}
by using $\varpi_{00}{*}\overline{\varpi}_{00}=0$ proved in 
\cite{ommy6}, but here we repeat the calculation.  
$$
\int_{-\pi}^{\pi}e_*^{(s{+}i\sigma)\frac{1}{i\h}u{*}v}d\sigma{*}
\int_{-\pi}^{\pi}e_*^{(s'{+}i\sigma')\frac{1}{i\h}v{*}u}d\sigma'
=
\int_{-\pi}^{\pi}\int_{-\pi}^{\pi}
e_*^{(s{+}i\sigma)\frac{1}{i\h}u{*}v{+}(s'{+}i\sigma')\frac{1}{i\h}v{*}u}
d\sigma d\sigma'
$$
can be defined always to give $0$, for by using $\frac{1}{i\h}u{*}v=\frac{1}{i\h}u{\ctt}v{-}\frac{1}{2}$,
and $\frac{1}{i\h}v{*}u=\frac{1}{i\h}u{\ctt}v{+}\frac{1}{2}$, 
the change of variables gives
$$
\int_{-\pi}^{\pi}e^{+i\sigma}d\sigma\int_{-\pi}^{\pi}
e^{\frac{1}{2}(s'{-}s{-}i\tau}e_*^{(s{+}s'{+}i\tau)\frac{1}{i\h}u{\ctt}v}d\tau=0.
$$
Similarly one can prove 
$E_{k,k}(K){*}D_{n,n}(K)=0$ by direct calculations, but 
computing  
$$
D_{n,n}(K){*_{_K}}{:}\frac{1}{i\h}u{\ctt}v{:}_{_K}{*_{_K}}E_{k,k}(K)
$$ 
two ways $(a{*}b){*}c=a{*}(b{*}c)$. We see $D_{n,n}(K){*}E_{k,k}(K){=}0$.
That is, identities above are requested to protect associativity.

\section{Fourier expansion of ${:}e_*^{(s{+}it)\frac{1}{i\h}u{\ctt}v}{:}_{_K}$}
Recall that the periodicity of ${:}e_*^{(s{+}it)\frac{1}{i\h}u{\ctt}v}{:}_{_K}$
w.r.t. $t$ depends on  $s$, but it is $4\pi$-periodic. 
Every $f(\theta)$ $4\pi$-periodic function is written as the sum of
a $2\pi$-periodic and a alternating $2\pi$-periodic functions: 
$$
f(\theta)=f_0(\theta){+}f_{-}(\theta),\quad 
f_0(\theta){=}\frac{1}{2}(f(\theta){+}f(\theta{+}2\pi)),\quad 
f_{-}(\theta){=}\frac{1}{2}(f(\theta){-}f(\theta{+}2\pi)).
$$ 

The Fourier series of $f(\theta)$ is given as 
$$
\begin{aligned}
f(\theta)=&\sum_{n}\frac{1}{4\pi}\int_0^{4\pi}f(t)e^{-\frac{1}{2}in t}dt
e^{i\frac{n}{2}\theta}\\
=&\sum_{n}\int_0^{2\pi}f_0(t)e^{-int}dte^{in\theta}+
\sum_{n}\frac{1}{2\pi}\int_0^{2\pi}f_{-}(t)
e^{-i(n{+}\frac{1}{2})t}dte^{i(n{+}\frac{1}{2})\theta}.
\end{aligned}
$$
As the periodicity of $e_*^{(s{+}it)\frac{1}{i\h}(u{\ctt}v}$
w.r.t. $t$ depends on $s$, we have to use Fourier basis 
depending on $s$. We denote the Fourier expansion 
\begin{equation}\label{Fourier}
{:}e_*^{(s{+}it)\frac{1}{i\h}u{\ctt}v}{:}_{_{K}}=
\left\{
\begin{matrix}
\medskip
\sum_{k=0}^{\infty}\tilde E_{k}(K)e^{(s{+}it)(k{+}\frac{1}{2})},& \quad s<a\\
\medskip
\sum_{n=-\infty}^{\infty}\tilde{D}_{n}(K)e^{(s{+}it)n},&\quad a<s<b \\
\medskip
\sum_{k=0}^{\infty}\tilde{E}_{-k}(K)e^{-(s{+}it)(k{+}\frac{1}{2})},&
\quad b<s\\
\end{matrix}
\right.
\end{equation}
where 
$$
\tilde E_{k}(K)=\frac{1}{2\pi}\int_{0}^{2\pi}
{:}e_*^{(s{+}it)\frac{1}{i\h}u{\ctt}v}{:}_{_{K}}e^{-(s{+}it)(k{+}\frac{1}{2})}dt, 
\quad 
\tilde{D}_{n}(K)=\frac{1}{2\pi}\int_{0}^{2\pi}
{:}e_*^{(s{+}it)\frac{1}{i\h}u{\ctt}v}{:}_{_{K}}e^{-(s{+}it)n}dt
$$
$$
\tilde E_{-k}(K)=\frac{1}{2\pi}\int_{0}^{2\pi}
{:}e_*^{(s{+}it)\frac{1}{i\h}u{\ctt}v}{:}_{_{K}}e^{-(s{+}it)(-k{-}\frac{1}{2})}dt,
$$
Cauchy's integral theorem  shows that these are independent of $s$ whenever
 $s<a$, $a<s<b$ and $b<s$ respectively. 
 
Thus, taking $s\to-\infty$ (resp. $s\to\infty$) we see that 
\begin{equation}\label{EE}
\tilde E_{k}(K)={E}_{k,k}(K),\quad \tilde
E_{-k}(K)=\overline{E}_{k,k}(K),\quad 
\text{if  }\,\, a{<}0{<}b, \text{then }\,\, {\tilde D}_{n}(K)=D_{n,n}(K). 
\end{equation}

However, it should be noted that these identities do not directly imply the idempotency
$$
\tilde E_{k}(K){*}\tilde E_{k}(K)=\tilde E_{k}(K),\quad 
\tilde E_{-k}(K){*}\tilde E_{-k}(K)=\tilde E_{-k}(K),\quad
{\tilde D}_{n}(K){*}{\tilde D}_{n}(K)={\tilde D}_{n}(K). 
$$

\medskip
\noindent 
{\bf Note} $\tilde{D}_{n}(K)$ was denoted by $\tilde{D}_{n,n}(K)$ in the previous
note \cite{ommy6}. As it is confusing, we changed the notation as above. 

\medskip
It is clear that $\tilde{D}_{n}(K)={D}_{n,n}(K)$ if $a<0<b$. In
general $\tilde{D}_{n}(K)$ may not be idempotent. However, 
$\tilde{D}_{n}(K)$ has an almost idempotent property as follows:
\begin{lem}
For $a<s, s'<b$, if there are $s, s'$ such that $a<s{+}s'<b$ then 
$$
\frac{1}{2\pi}\int_{0}^{2\pi}
{:}e_*^{(s{+}it)\frac{1}{i\h}u{\ctt}v}{:}_{_K}e^{(s{+}it)n}dt{*_{_K}}
\frac{1}{2\pi}\int_{0}^{2\pi}
{:}e_*^{(s'{+}it)\frac{1}{i\h}u{\ctt}v}{:}_{_K}e^{(s{+}it)n}dt{=}
\frac{1}{2\pi}\int_{0}^{2\pi}
{:}e_*^{(s{+}it)\frac{1}{i\h}u{\ctt}v}{:}_{_K}e^{(s{+}it)n}dt.
$$
If such $s, s'$ can not be selected, then 
 $\int_{0}^{2\pi}
{:}e_*^{(s{+}it)\frac{1}{i\h}u{\ctt}v}{:}_{_K}e^{(s{+}it)n}dt{*}
\int_{0}^{2\pi}
{:}e_*^{(s'{+}it)\frac{1}{i\h}u{\ctt}v}{:}_{_K}e^{(s{+}it)n}dt$ is not welldefined.
\end{lem}

\noindent
{\bf Note}\,\,\,Suppose $s, s'\in I_{\ctt}(K)$ but
$s{+}s'\not\in I_{\ctt}(K)$.  By the $2\pi$-periodicity, we have 
$$
\frac{1}{2\pi}\int_{0}^{2\pi}
{:}e_*^{(\sigma{+}it)\frac{1}{i\h}u{\ctt}v}{:}_{_K}e^{(s{+}it)n}dt=
\frac{1}{4\pi}\int_{0}^{4\pi}
{:}e_*^{(\sigma{+}it)\frac{1}{i\h}u{\ctt}v}{:}_{_K}e^{(s{+}it)n}dt,
\text{for}\,\,\,\sigma=s, s.'
$$

However, the product $e_*^{(s{+}s'{+}i(t{+}t'))\frac{1}{i\h}u{\ctt}v}$
is alternating $2\pi$-periodic hence 
$$
(\frac{1}{4\pi})^2\int_{0}^{4\pi}\int_{0}^{4\pi}e_*^{(s{+}s'{+}i(t{+}t'))\frac{1}{i\h}u{\ctt}v}dtdt'=0.
$$
But
$$
(\frac{1}{2\pi})^2\int_{0}^{2\pi}\int_{0}^{2\pi}e_*^{(s{+}s'{+}i(t{+}t'))\frac{1}{i\h}u{\ctt}v}dtdt'=
\frac{1}{2\pi}\int_{0}^{2\pi}e_*^{(s{+}s'{+}it)\frac{1}{i\h}u{\ctt}v}dt\not=0.
$$
This sounds strange, for $a=b$ does not give $a^{2}{=}b^{2}$. Such
strange phenomena caused by the discontinuity of 
$\int_{0}^{4\pi}e_*^{(s{+}s'{+}it)\frac{1}{i\h}u{\ctt}v}dt$.

\bigskip
To avoid the possible confusion, it is better to use notations similar
to \eqref{vectdist}, \eqref{vectdist2}:  Setting
$$
V_{n}^{(+)}(s;K)=
\frac{1}{4\pi}{:}\int_{-2\pi}^{2\pi}
e_*^{(s{+}it)\frac{1}{i\h}u{\ctt}v}e^{-(s{+}it)(n{+}\frac{1}{2})}dt{:}_{_{K}},\quad
V_{n}^{(0)}(s;K)=
\frac{1}{4\pi}{:}\int_{-2\pi}^{2\pi}
e_*^{(s{+}it)\frac{1}{i\h}u{\ctt}v}e^{-(s{+}it)n}dt{:}_{_{K}},
$$
we denote 
\begin{equation}\label{vectdist3}
\begin{aligned}
&(\frac{1}{i\h}u{\ctt}v){*}V_{n}^{(+)}(s;*)={-}\delta(s{-}a)\tilde E_{n}{+}\delta(s{-}b)\tilde E_{-n}\\
&(\frac{1}{i\h}u{\ctt}v){*}V_{n}^{(0)}(s;*)=\delta(s{-}a)\tilde D_{n}{-}\delta(s{-}b)\tilde D_{n}.
\end{aligned}
\end{equation}

\subsection{Element given by Fourier series}

Suppose $c_n$ is given as Fourier coefficients of a Schwartz 
distribution $h(z)$ defined on $|z|=1$.
Using its Fourier series we define three functions:  
$$
h_0(t){=}\!\sum_{n=\infty}^{\infty}c_ne^{itn},\quad 
h_+(t)=e^{{-}\frac{1}{2}it}h(t){=}\!\sum_{n=\infty}^{\infty}c_ne^{it(n{-}\frac{1}{2})},
\quad 
h_-(t)=e^{\frac{1}{2}it}h(t){=}\!\sum_{n=\infty}^{\infty}c_ne^{it(n{+}\frac{1}{2})}.
$$
$h_0(t)$ is $2\pi$-periodic in w.r.t. $t$, and others are
$2\pi$-alternating periodic. 
Hence the pairing with 
${:}e_*^{(s{+}it)\frac{1}{i\h}u{\ctt}v}{:}_{_K}$ defined below give
respectively an element of $H\!ol({\mathbb C}^2)$:

\bigskip
\noindent
(1) If $a<s<b$, then 
$$
h_{0*}(s,\frac{1}{i\h}u{\ctt}v)=\frac{1}{2\pi}
\int_{0}^{2\pi}h(t){:}e_*^{(s{+}it)\frac{1}{i\h}u{\ctt}v}{:}_{_K}dt{=}
 \sum_{n=-\infty}^{\infty}c_{-n}e^{sn}\tilde{D}_{n}(K).
$$ 

This is  an element of $H\!ol({\mathbb C}^2)$.

\medskip
\noindent
(2) If $s<a$, then 
$$
\begin{aligned} 
h_{+*}(s,\frac{1}{i\h}u{\ctt}v)&{=}
\frac{1}{2\pi}\int_{0}^{2\pi}
e^{-\frac{1}{2}it}h(t){:}e_*^{(s{+}it)\frac{1}{i\h}u{\ctt}v}{:}_{_K}dt\\
&{=}
\frac{1}{2\pi}\int_{0}^{2\pi}\sum_{n\in{\mathbb Z}}c_{n}
e^{it(n{-}\frac{1}{2})}\sum_{k\geq 0}E_{k,k}(K)e^{(s{+}it)(k{+}\frac{1}{2})}dt{=}
\sum_{n=0}^{\infty}c_{-n}e^{s(n{+}\frac{1}{2})}E_{n,n}(K)
\end{aligned}
$$ 
as ${:}e_*^{(s{+}it)\frac{1}{i\h}u{\ctt}v}{:}_{_K}$ have only positive
Fourier coefficients. For $s<a$,
$\sum_{n=0}^{\infty}c_{-n}e^{s(n{+}\frac{1}{2})}E_{n,n}(K)$ is an element of $H\!ol({\mathbb C}^2)$,
but each $c_{-n}e^{s(n{+}\frac{1}{2})}E_{n,n}(K)$ is an element of
$H\!ol({\mathbb C}^2)$ for any $s$. Note also that
$\frac{1}{i\h}u{\ctt}v{-}\frac{1}{2}=\frac{1}{i\h}u{*}v$ to understand
the meaning of $\frac{1}{2}$. 

\medskip
\noindent
(3) If $b<s$, then 
$$
\begin{aligned}
\frac{1}{2\pi}\int_{0}^{2\pi}e^{\frac{1}{2}it}h(t){:}e_*^{(s{+}it)\frac{1}{i\h}u{\ctt}v}{:}_{_K}dt
{=}&
\frac{1}{2\pi}\int_{0}^{2\pi}\sum_{n\in{\mathbb Z}}c_{n}
e^{it(n{+}\frac{1}{2})}\sum_{k\geq 0}\overline{E}_{k,k}(K)e^{{-}(s{+}it)(k{+}\frac{1}{2})}dt\\
{=}&
\sum_{n=0}^{\infty}c_{n}e^{-s(n{+}\frac{1}{2})}\overline{E}_{n,n}(K)
\end{aligned}
$$
as 
$e^{\frac{1}{2}(s{+}it)}{:}e_*^{(s{+}it)\frac{1}{i\h}u{\ctt}v}{:}_{_K}$ 
is holomorphic on the on the outside of the
disk $|e^{s{+}it}|{\geq}e^{b}$ having only negative Fourier 
coefficients.  
$\sum_{n=0}^{\infty}c_{n}e^{-s(n{+}\frac{1}{2})}\overline{E}_{n,n}(K)$ 
is an element of $H\!ol({\mathbb C}^2)$, but each 
$c_{n}e^{-s(n{+}\frac{1}{2})}\overline{E}_{n,n}(K)$ is an element of
$H\!ol({\mathbb C}^2)$ for any $s$.

Then, these converge 
for $s<a$, $a<s<b$ and $b<s$  respectively in
$C^{\infty}(S^1,H\!ol({\mathbb C}^2))$ 
$($$H\!ol({\mathbb C}^2)$-valued smooth 
functions on $S^1$ with $C^\infty$-topology$)$.

Summarizing these we have 
\begin{thm}\label{Nicecoeff}
Let $\{c_n\}_{n\in{\mathbb Z}}$ be Fourier coefficients of a Schwartz
distribution on $S^1$. Then in generic ordered expression,
$\sum_{n=-\infty}^{\infty}c_{-n}e^{sn}\tilde{D}_{n}(K)$,
$\sum_{n=0}^{\infty}c_{-n}e^{s(n{+}\frac{1}{2})}E_{n,n}(K)$,
$\sum_{n=0}^{\infty}c_{n}e^{{-}s(n{+}\frac{1}{2})}\overline{E}_{n,n}(K)$ 
are elements of $H\!ol({\mathbb C}^2)$ respectively for 
$a<s<b$, $s<a$ and $b<s$.
\end{thm}
It is remarkable that there is no singular point in the diagonal
matrix expressions.

Applying Theorem\ref{Nicecoeff} to a function 
$h(z,t)=\sum_{n=-\infty}^{\infty}\frac{1}{(z{+}\frac{1}{2})_{-n}}e^{itn}$ 
depending on $z$, we see for instance that 
$$
\sum_{n=0}^{\infty}\frac{1}{(z{+}\frac{1}{2})_{n}}e^{s(n{+}\frac{1}{2})}E_{n,n}(K)
$$
converges to an element of $H\!ol({\mathbb C}^2)$, 
where $(a)_n=a(a{+}1)\cdots(a{+}n{-}1)$, $(a)_0=1$ and 
$(a)_{-n}=(a{-}1)(a{-}2)\cdots(a{-}n)$.

\subsection{Analytic continuation  of  $\delta_*(iz{+}\frac{1}{\h}u{\ctt}v)$}

In this section, we show that $\delta_*(iz{+}\frac{1}{\h}u{\ctt}v)$
behaves like an idempotent element in the calculation based on the
complex integral and residue calculus.
It is known that $\delta_*(iz{+}\frac{1}{\h}u{\ctt}v)$ is given by the
difference of two different inverses:
$$
\delta_*(iz{+}\frac{1}{\h}u{\ctt}v)
=(z{+}\frac{1}{i\h}u{\ctt}v)^{-1}_{*+}-(z{+}\frac{1}{i\h}u{\ctt}v)^{-1}_{*-}
$$ 
which is holomorphic on the domain  $|{\rm{Re}}z|<\frac{1}{2}$. 
By making analytic continuations of inverses, we have seen in
\cite{ommy6} the next 
\begin{thm}\label{deltacont}
$\delta_*(iz{+}\frac{1}{\h}u{\ctt}v)$ is analytically 
continued on the space 
${\mathbb C}{\setminus}({\mathbb Z}+\frac{1}{2})$ 
as an $H{\!o}l({\mathbb C}^2)$-valued 
 holomorphic functions of $z$ with simple poles. 

The residues of ${:}\delta_*(iz{+}\frac{1}{\h}u{\ctt}v){:}_{_K}$ 
at ${-}(n{+}\frac{1}{2})$, $(n{+}\frac{1}{2})$ 
are $E_{n,n}(K)$, ${-}\overline{E}_{n,n}(K)$ respectively.
\end{thm}

Theorem\,\ref{deltacont} can be proved directly by taking integral of \eqref{Fourier} as
follows:     
$$
{:}\delta_*(iz{+}\frac{1}{\h}u{\ctt}v){:}_{_K}=
\sum_{k=0}^{\infty}\frac{e^{a(z{+}k{+}\frac{1}{2})}}
{z{+}(k{+}\frac{1}{2})}{\tilde E}_{k}(K){+}
\sum_{n=-\infty}^{\infty}\frac{e^{b(z{+}n)}{-}e^{a(z{+}n)}}{z{+}n}{\tilde D}_n(K){-}
\sum_{k=0}^{\infty}\frac{e^{b(z{-}(k{+}\frac{1}{2})}}{z{-}(k{+}\frac{1}{2})}{\tilde E}_{-k}(K)
$$
Note that 
$\frac{e^{b(z{+}n)}{-}e^{a(z{+}n)}}{z{+}n}{\tilde D}_n(K)$ is not
singular at $z=-n$, and the singular points and the residues are
calculated by the terms 
${\tilde E}_{k}(K)$, $k\in {\mathbb Z}$.  
\medskip
Thus, analytically continued $\delta_*(\zeta{+}\frac{1}{\h}u{\ctt}v)$ 
may be viewed in generic ordered expression $K$ that 
$$
{:}\delta_*(\zeta{+}\frac{1}{\h}u{\ctt}v){:}_{_K}=
\sum_{n=0}^{N}\frac{i}{\zeta{+}i(n{+}\frac{1}{2})}E_{n,n}(K)
 -\sum_{n=0}^{N}
\frac{i}{\zeta{-}i(n{+}\frac{1}{2})}\overline{E}_{n,n}(K)+\Phi_K^{N}(\zeta), 
$$
where $\Phi_K^{N}(\zeta)$ is holomorphic on
$|{\rm{Im}}\,\zeta|<N{+}\frac{1}{2}$.
Theorem\,\ref{deltacont} gives also the following: 
\begin{prop}\label{Matrix}
For every holomorphic function $f(z)$ defined on a 
simply connected domain $D$, the following equality holds for
generic ordered expression:
For every smooth simple closed curve $C$ in $D$:  
\begin{equation}\label{compinteg}
\frac{1}{2\pi i}\int_C f(w)
\delta_*(z{-}w{+}\frac{1}{\h}u{\ctt}v)dw 
=\sum_{n}f(z{+}i(n{+}\frac{1}{2}))E_{n,n}{-} 
 \sum_{n}f(z{-}i(n{+}\frac{1}{2}))\overline{E}_{n,n}
\end{equation}
where the each summation runs through all integers 
$n\geq 0$ such that $-i(n{+}\frac{1}{2})$ and 
$i(n{+}\frac{1}{2})$ are inside of $C$ respectively. 
\end{prop}

Thus, we define  
\begin{equation}\label{formaldist}
f_*^{C}(z{+}\frac{1}{\h}u{\ctt}v)= 
\frac{1}{2\pi i}\int_C f(w) 
\delta_*(z{-}w{+}\frac{1}{\h}u{\ctt}v)dw 
\end{equation}
for every holomorphic function $f$ and  
$$
1_*^{C}=\frac{1}{2\pi i}\int_C 
\delta_*(z{-}w{+}\frac{1}{i\h}u{\ctt}v)dw=
\sum_k{E}_{k,k}(K)-\sum_k\overline{E}_{k,k}(K).
$$

Now recalling that    
$E_{n,n}{*}\overline{E}_{k,k}=0=
\overline{E}_{n,n}{*}{E}_{k,k}$, we have the idempotency  
$$
\frac{1}{2\pi i}\int_C 
\delta_*(z{-}w{+}\frac{1}{i\h}u{\ctt}v)dw{*}
\frac{1}{2\pi i}\int_C 
\delta_*(z{-}w{+}\frac{1}{i\h}u{\ctt}v)dw{=}
\frac{1}{2\pi i}\int_C 
\delta_*(z{-}w{+}\frac{1}{i\h}u{\ctt}v)dw.
$$
Hence we have  
\begin{equation}\label{compose}
f_*^{C}(z{+}\frac{1}{\h}u{\ctt}v){*} 
g_*^{C}(z{+}\frac{1}{\h}u{\ctt}v)= 
\frac{1}{2\pi i}\int_C f(w)g(w) 
\delta_*(z{-}w{+}\frac{1}{\h}u{\ctt}v)dw. 
\end{equation}
This is confirmed also by the resolvent calculus as follows: 
\begin{lem}\label{orthdelta}
If $w\not=w'$, then 
$$
\delta_{*}(z{-}w{+}\frac{1}{\h}u{\ctt}v){*}\delta_{*}(z{-}w'{+}\frac{1}{\h}u{\ctt}v)=0.
$$
\end{lem}

\noindent
{\bf Proof}\,\, Let 
$$
\begin{aligned}
\delta_{*}(z{-}w{+}\frac{1}{\h}u{\ctt}v)
=&(z{-}w{+}\frac{1}{\h}u{\ctt}v)^{-1}_{*+}{-}(z{-}w{+}\frac{1}{\h}u{\ctt}v)^{-1}_{*-},\\
\delta_{*}(z{-}w'{+}\frac{1}{\h}u{\ctt}v)
=&(z{-}w'{+}\frac{1}{\h}u{\ctt}v)^{-1}_{*+}{-}(z{-}w'{+}\frac{1}{\h}u{\ctt}v)^{-1}_{*-}.
\end{aligned}
$$
In the resolvent calculus, the product of the inverse is given by 
$$ 
\frac{1}{w'{-}w}
\Big((z{-}w{+}\frac{1}{\h}u{\ctt}v)^{-1}_{*\pm}{-}(z{-}w'{+}\frac{1}{\h}u{\ctt}v)^{-1}_{*\pm}
\Big).
$$
Hence the result follows by a direct computation. \hfill $\Box$

Recalling \eqref{compose}, one may write this by 
\begin{equation}
\delta_{*}(z{-}w{+}\frac{1}{\h}u{\ctt}v){*}\delta_{*}(z{-}w'{+}\frac{1}{\h}u{\ctt}v)=
\delta(w'{-}w)\delta_{*}(z{-}w{+}\frac{1}{\h}u{\ctt}v), \quad cf. \eqref{deltaidem}.
\end{equation}
Here, note that $\delta(w'{-}w)$ is the delta function of complex 
variables regarded as a {\it formal distribution}.

\noindent
{\bf Note}\,\,Formal distribution is the extended notion of
distribution, using Laurent polynomials as test functions and 
residues as the integrations. This is convenient to 
algebraic calculations.

\bigskip
Since 
$$
(\frac{1}{i\h}u{\ctt}v){*}E_{n,n}(K)=(n{+}\frac{1}{2})E_{n,n}(K), \quad 
(\frac{1}{i\h}u{\ctt}v){*}\overline{E}_{n,n}(K)=-(n{+}\frac{1}{2})\overline{E}_{n,n}(K),
$$
we have also that 
$$
\begin{aligned}
{:}(\frac{1}{i\h}u{\ctt}v){*}\delta_*(z+\frac{1}{\h}u{\ctt}v){:}_{_K}=&
\sum_{n=0}^{\infty}\frac{i(n{+}\frac{1}{2})}{z{+}i(n{+}\frac{1}{2})}E_{n,n}(K){-}
\sum_{n=0}^{\infty}
\frac{i(n{+}\frac{1}{2})}{z{-}i(n{+}\frac{1}{2})}\overline{E}_{n,n}(K)+(\frac{1}{i\h}u{\ctt}v){*}\Phi_K(z)\\ 
=&-z{:}\delta_*(z+\frac{1}{\h}u{\ctt}v){:}_{_K} +1^{C}{+}(\frac{1}{i\h}u{\ctt}v){*}\Phi_K(z).
\end{aligned}
$$
Since the extended $\delta_{*}(z{-}w{+}\frac{1}{\h}u{\ctt}v)$ is given
by the difference of two different inverses, we have 
$$
(z{+}\frac{1}{\h}u{\ctt}v){*}\frac{1}{2\pi i}\int_{C}\delta_{*}(z{-}w{+}\frac{1}{\h}u{\ctt}v)dw{=}
\frac{1}{2\pi i}\int_{C}w\delta_{*}(z{-}w{+}\frac{1}{\h}u{\ctt}v)dw{=}
(z{+}\frac{1}{\h}u{\ctt}v)_*^{C}.
$$

\noindent
{\bf Note}\,\,The procedure of taking $\frac{1}{2\pi i}\int_{C}dw$ is
useful when we concern only on residues.

\subsubsection{Matrix elements}\label{matelm01} 
The proof of Lemma\,\ref{orthdelta} extends to give  
\begin{lem}\label{orthdelta22}
If $w{-}w'$ is not an integer, then for every integer $n\geq 0$ 
$$
\delta_{*}(z{-}w{+}\frac{1}{\h}u{\ctt}v){*}u^n{*}
\delta_{*}(z{-}w'{+}\frac{1}{\h}u{\ctt}v)=0=
\delta_{*}(z{-}w{+}\frac{1}{\h}u{\ctt}v){*}v^n{*}
\delta_{*}(z{-}w'{+}\frac{1}{\h}u{\ctt}v).
$$
\end{lem}

\noindent
{\bf Proof}\,\,Recall the formula 
\begin{equation}\label{factrial}
(\frac{1}{i\h})^nu^n{*}v^n{=}(\frac{1}{i\h}u{\ctt}v{-}\frac{1}{2}){*}
(\frac{1}{i\h}u{\ctt}v{-}\frac{3}{2}){*}\cdots{*}(\frac{1}{i\h}u{\ctt}v{-}\frac{2n-1}{2}).
\end{equation}
Using $\int_0^{\infty}e_*^{t(\frac{1}{i\h}u{\ctt}v{-}\alpha)}dt$, 
one can obtain an inverse $(u^n{*}v^n)_{*-}^{-1}$.

Under the independent $\pm$ sign, we consider  
$(z{-}w{+}\frac{1}{i\h}u{\ctt}v)_{*\pm}^{-1}{*}(u^n{*}(z{-}w'{+}\frac{1}{i\h}u{\ctt}v)_{*\pm}^{-1}{*}v^n)$.
As $v^n{*}u^n$ commutes with $(z{-}w'{+}\frac{1}{\h}u{\ctt}v)$, we see this is an inverse of 
$$
(u^n{*}v^n)_{*-}^{-2}(u^n{*}(z{-}w'{+}\frac{1}{i\h}u{\ctt}v){*}v^n){*}(z{-}w{+}\frac{1}{i\h}u{\ctt}v)
=
(u^n{*}v^n)_{*-}^{-1}(z{-}w'{-}n{+}\frac{1}{i\h}u{\ctt}v){*}(z{-}w{+}\frac{1}{i\h}u{\ctt}v)
$$ 
On the other hand, the resolvent calculus gives that this inverse is 
obtained by 
$$
\frac{1}{w'{-}w{+}n}\Big(u^n{*}(z{-}w'{+}\frac{1}{i\h}u{\ctt}v)_{*\pm}^{-1}{*}v^n{-}
u^n{*}v^n{*}(z{-}w{+}\frac{1}{i\h}u{\ctt}v)_{*\pm}^{-1}\Big),
$$
for 
$$
\begin{aligned}
&(u^n{*}v^n)_{*-}^{-2}(u^n{*}(z{-}w'{+}\frac{1}{i\h}u{\ctt}v){*}v^n){*}(z{-}w{+}\frac{1}{i\h}u{\ctt}v)\\
&\qquad\qquad\qquad{*}\Big(u^n{*}(z{-}w'{+}\frac{1}{i\h}u{\ctt}v)_{*\pm}^{-1}{*}v^n{-}
     u^n{*}v^n{*}(z{-}w{+}\frac{1}{i\h}u{\ctt}v)_{*\pm}^{-1}\Big)\\
&=(z{-}w{+}\frac{1}{i\h}u{\ctt}v){-}(z{-}w'{-}n{+}\frac{1}{i\h}u{\ctt}v){=}w'{+}n{-}w
\end{aligned}
$$
By the similar calculations as in Lemma\,\ref{orthdelta}, we see 
$$
\delta_{*}(z{-}w{+}\frac{1}{\h}u{\ctt}v){*}
(u^n{*}\delta_{*}(z{-}w'{+}\frac{1}{\h}u{\ctt}v){*}v^n){=}0.
$$
As $v_*^{\ctt}=u{*}\int_{-\infty}^{0}e_*^{s\frac{1}{i\h}v{*}u}ds$ is a
left inverse of $v$ i.e. $v{*}v_*^{\ctt}=1$, $v_*^{\ctt}{*}v{=}1{-}\varpi_{00}$,
we have 
$$
\delta_{*}(z{-}w{+}\frac{1}{\h}u{\ctt}v){*}
u^n{*}\delta_{*}(z{-}w'{+}\frac{1}{\h}u{\ctt}v)=
\delta_{*}(z{-}w{+}\frac{1}{\h}u{\ctt}v){*}
(u^n{*}\delta_{*}(z{-}w'{+}\frac{1}{\h}u{\ctt}v){*}v^n){*}(v^{\ctt}_{*})^n{=}0.
$$
${}$\hfill $\Box$

In what follows, we use the convention \eqref{pmconvention} and the
notations 
$$
(a)_n=a(a{+}1)\cdots(a{+}n{-1}),\quad (a)_0=1,\quad  
(a)_{-n}=(a{-}1)(a{-}2)\cdots(a{-}n).
$$
Note also that 
\begin{equation}\label{factrial55}
\begin{aligned}
&(\frac{1}{i\h})^nv^n{*}u^n{=}(\frac{1}{i\h}u{\ctt}v{+}\frac{1}{2}){*}
(\frac{1}{i\h}u{\ctt}v{+}\frac{3}{2}){*}\cdots{*}(\frac{1}{i\h}u{\ctt}v{+}\frac{2n-1}{2})
{=}(\frac{1}{i\h}u{\ctt}v{+}\frac{1}{2})_{n*},\\
&(\frac{1}{i\h})^nu^n{*}v^n{=}(\frac{1}{i\h}u{\ctt}v{-}\frac{1}{2}){*}
(\frac{1}{i\h}u{\ctt}v{-}\frac{3}{2}){*}\cdots{*}(\frac{1}{i\h}u{\ctt}v{-}\frac{2n-1}{2})
{=}(\frac{1}{i\h}u{\ctt}v{+}\frac{1}{2})_{-n*}.
\end{aligned}
\end{equation}

As $\delta_{*}(z{-}w{+}\frac{1}{\h}u{\ctt}v)$ is a difference of two
different inverses, we see 
$$
(\frac{1}{i\h})^n{\hat\zeta}^n{*}\zeta^n{*}\delta_{*}(z{-}w{+}\frac{1}{\h}u{\ctt}v)=
(i(w{-}z){+}\frac{1}{2})_{n}\delta_{*}(z{-}w{+}\frac{1}{\h}u{\ctt}v),\quad
n\in{\mathbb Z}.
$$

Set
\begin{equation}\label{frahddd}
{\mathfrak D}_{p,q}(z{+}\frac{1}{\h}u{\ctt}v)=
\frac{1}{2\pi i}
\int_{C}
\frac{1}{\sqrt{(i\h)^{p{+}q}(i(w{-}z){+}\frac{1}{2})_{p}(i(w{-}z){+}\frac{1}{2})_{q}}}
\zeta^p{*}\delta_{*}(z{-}w{+}\frac{1}{\h}u{\ctt}v){*}{\hat\zeta}^qdw,
\end{equation}
where $C$ is a closed curve.
It is not hard to see that 
${\mathfrak D}_{p,q}(z{+}\frac{1}{\h}u{\ctt}v)$ form matrix elements. 
$$
{\mathfrak D}_{p,q}(z{+}\frac{1}{\h}u{\ctt}v){*}{\mathfrak D}_{r,s}(z'{+}\frac{1}{\h}u{\ctt}v)
{=}\delta_{q,r}\delta(z{-}z'){\mathfrak D}_{p,s}(z{+}\frac{1}{\h}u{\ctt}v).
$$ 

\section{Integrals along real axis}

So far we have mainly concerned with the Fourier expansion of 
$e_*^{(s{+}it)\frac{1}{i\h}u{\ctt}v}$ w.r.t. the variable $t$, and 
we have seen there is a big difference between differentiations by real
variables and by pure imaginary variables. 
Another remarkable feature of 
$e_*^{(s{+}it)\frac{1}{i\h}u{\ctt}v}$ is that   
if $|{\rm{Re}}\,z|<1/2$ then 
\begin{equation}\label{deltaintg}
\int_{-\infty}^{\infty}e^{s(z{+}\frac{1}{i\h}u{\ctt}v+i\xi)}ds=
\int_{-\infty}^{\infty}e^{-is(iz{+}\frac{1}{\h}u{\ctt}v{-}\xi)}ds=
\delta_*(iz{+}\frac{1}{\h}u{\ctt}v{-}\xi),\quad |{\rm{Re}}\,z|<1/2.
\end{equation}
is rapidly decreasing w.r.t. $\xi\in{\mathbb R}$. In this section we
discuss what this property produces.
 
For every tempered distribution $f(\xi)$ on $\mathbb R$, we define 
\begin{equation}\label{tempdist}
f_*(z{+}\frac{1}{\h}u{\ctt}v)=
\frac{1}{2\pi}\int_{-\infty}^{\infty}f(\xi)\delta_*(iz{-}\xi{+}\frac{1}{\h}u{\ctt}v)d\xi.
\end{equation}
By \eqref{deltaidem}, if the pointwise product $f(\xi)g(\xi)$ is defined as a tempered
distribution, then we have a formula similar to \eqref{compose}
\begin{equation}\label{comose2}
f_*(z{+}\frac{1}{\h}u{\ctt}v){*}g_*(z{+}\frac{1}{\h}u{\ctt}v)=
\int_{-\infty}^{\infty}f(\xi)g(\xi)\delta_*(iz{-}\xi{+}\frac{1}{\h}u{\ctt}v)d\xi.
\end{equation}
For the characteristic function $\chi_{U}(\xi)$ of an open subset 
$\chi_{U*}(z{+}\frac{1}{\h}u{\ctt}v)$ is an idempotent element. 

Let $Y_{\pm}(t)$ be the characteristic function of $(0,\infty)$,
$(-\infty,0)$.   
Let $\hat{Y}_{\pm}(xi)$ be the Fourier transform of $Y_{\pm}(t)$.  
Then, we have 
\begin{equation}\label{Heaviside}
(z{+}\frac{1}{\h}u{\ctt}v)_{*\pm}^{-1}=
\frac{1}{2\pi}\int_{\mathbb R}\hat{Y}_{\pm}(xi)\delta_*(iz{-}\xi{+}\frac{1}{\h}u{\ctt}v)d\xi.
\end{equation}

Similarly, we have for every $a\in{\mathbb R}$ 
$$
e_*^{ai(iz{+}\frac{1}{\h}u{\ctt}v)}=
\frac{1}{2\pi}\int_{\mathbb R}e^{ai\xi}\delta_*(iz{-}\xi{+}\frac{1}{\h}u{\ctt}v)d\xi.
$$
Since $\lim_{n\to\infty}(1{+}\frac{aiw}{n})^n{=}e^{aiw}$ as tempered distribution, we have 
in generic ordered expression that
$$
\lim_{n\to\infty}\Big(1{+}\frac{ai(iz{+}\frac{1}{\h}u{\ctt}v)}{n}\Big)_*^n{=}
e_*^{ai(iz{+}\frac{1}{\h}u{\ctt}v)}.
$$

\medskip
\noindent
{\bf Extended properties of
  $\delta_*(iz{+}\frac{1}{\h}u{\ctt}v{-}\xi)$ and matrix elements}

As in \S\,\ref{matelm01}, matrix elements are produced by the
integrals along the real axis. Minding \eqref{frahddd}, we set  
$$
{\mathcal D}_{p,q}(z{+}\frac{1}{\h}u{\ctt}v)=
\frac{1}{2\pi}\int_{\mathbb R}
\frac{1}{\sqrt{(i\h)^{p{+}q}(\xi{-}iz{+}\frac{1}{2})_{p}(\xi{-}iz{+}\frac{1}{2})_{q}}}
\,\,\zeta^p{*}\delta_{*}(z{-}\xi{+}\frac{1}{\h}u{\ctt}v){*}{\hat\zeta}^qd\xi.
$$

Note first that Lemmas\,\ref{orthdelta}, \ref{orthdelta22} can be
applied to this case. Thus, we see 
$$
{\mathcal D}_{p,q}(z{+}\frac{1}{\h}u{\ctt}v){*}
{\mathcal D}_{r,s}(z'{+}\frac{1}{\h}u{\ctt}v){=}
\delta_{q,r}\delta(z{-}z'){\mathcal D}_{p,s}(z{+}\frac{1}{\h}u{\ctt}v)
$$

\subsection{Diagonal matrix calculations} 

Let $I_{\ctt}(K)=(a,b)$ be the exchanging interval for a generic expression 
parameter $K$. 
$(z{+}\frac{1}{i\h}u{\ctt}v)_{*{\pm}}^{-1}$ is given by 
$$
(z{+}\frac{1}{i\h}u{\ctt}v)_{*{+}}^{-1}=\int_{-\infty}^0e_*^{t(z{+}\frac{1}{i\h}u{\ctt}v)}dt,\,\,
({\rm{Re}}\,z>-\frac{1}{2}), 
\quad 
(z{+}\frac{1}{i\h}u{\ctt}v)_{*{-}}^{-1}=-\int_{0}^{\infty}e_*^{t(z{+}\frac{1}{i\h}u{\ctt}v)}dt\,\,
({\rm{Re}}\,z<\frac{1}{2}).
$$ 
It is easy to see that in $H\!ol({\mathbb C}^2)$
$$
{:}(z{+}\frac{1}{i\h}u{\ctt}v)_{*{+}}^{-1}{:}_{_K}=
{:}e_*^{-a(z{+}\frac{1}{i\h}u{\ctt}v)}{:}_{_K}{*_{_K}}
{:}\!\int_{-\infty}^ae_*^{t(z{+}\frac{1}{i\h}u{\ctt}v)}dt{:}_{_K}{=}
{:}e_*^{-a(z{+}\frac{1}{i\h}u{\ctt}v)}{:}_{_K}{*_{_K}}
\Big(\sum_{k=0}^{\infty}\frac{e^{a(z{+}n{+}\frac{1}{2})}}{z{+}n{+}\frac{1}{2}}E_{k,k}(K)\Big),
$$
but the next equality holds only if the right-multiplication operation 
$e_*^{-a(z{+}\frac{1}{i\h}u{\ctt}v)}{*}$ is continuous. 
However, if one concerns only with coefficients of diagonal matrix
elements, then 
$$
{=}\sum_{k=0}^{\infty}\frac{e^{a(z{+}k{+}\frac{1}{2})}}{z{+}k{+}\frac{1}{2}}
{:}e_*^{-a(z{+}\frac{1}{i\h}u{\ctt}v)}{:}_{_K}{*_{_K}}E_{k,k}(K){=}
\sum_{k=0}^{\infty}\frac{1}{z{+}k{+}\frac{1}{2}}E_{k,k}(K).
$$
Thus, we denote this equality as 
$$
{:}(z{+}\frac{1}{i\h}u{\ctt}v)^{-1}{:}_{_{E(K)mat}}{=}
\sum_{k=0}^{\infty}\frac{1}{z{+}k{+}\frac{1}{2}}E_{k,k}(K).
$$
If $K\in {\mathfrak K}_+$, 
$\sum_{k=0}^{\infty}\frac{1}{z{+}n{+}\frac{1}{2}}E_{k,k}(K)$ converges
in $H\!ol({\mathbb C}^2)$ as partial fractions. Hence 
${:}(z{+}\frac{1}{i\h}u{\ctt}v)^{-1}{:}_{_{E(K)mat}}$ may be viewed 
as the expression by the virtual expression parameter $K$ such that 
$I_{\ctt}(K)=(\infty,\infty)$. 
${E(K)mat}$-expression makes sense in $H\!ol({\mathbb C}^2)$, if 
$K\in {\mathfrak K}_+$.

\medskip
If $a<0<b$, then we have a hybrid expression 
$$
\begin{aligned}
{:}(z{+}\frac{1}{i\h}u{\ctt}v)_{*{+}}^{-1}{:}_{_K}&{=}
{:}\!\int_{-\infty}^ae_*^{t(z{+}\frac{1}{i\h}u{\ctt}v)}dt{:}_{_K}{+}
{:}\!\int_{a}^0e_*^{t(z{+}\frac{1}{i\h}u{\ctt}v)}dt{:}_{_K}\\
&{=}
\sum_{k=0}^{\infty}\frac{e^{a(z{+}\frac{1}{2}{+}k)}}{z{+}\frac{1}{2}{+}k}E_{k,k}(K)
{+}\sum_{n=-\infty}^{\infty}\frac{1{-}e^{a(z{+}n)}}{z{+}n}\tilde{D}_n(K).
\end{aligned}
$$
However if we take the limit $a\to -\infty$, $b\to \infty$, then the
second term diverges. Thus, 
the ${D}(K)mat$-expression ${:}(z{+}\frac{1}{i\h}u{\ctt}v)^{-1}{:}_{_{D(K)mat}}$
diverges. Similarly    
${:}(z{+}\frac{1}{i\h}u{\ctt}v)^{-1}{:}_{_{\overline{E}(K)mat}}$
diverges, but we see that  
if $K\in{\mathfrak K}_-$, then recalling \eqref{Fourier} we see that 
$$
{:}(z{-}\frac{1}{i\h}u{\ctt}v)^{-1}{:}_{_{\overline{E}(K)mat}}=
\sum_{k=0}^{\infty}\frac{1}{z{+}k{+}\frac{1}{2}}\overline{E}_{k,k}(K)
$$ 
converges in $H\!ol({\mathbb C}^2)$ as partial fractions. 
${\overline{E}(K)mat}$-expression makes sense in $H\!ol({\mathbb C}^2)$, if 
$K\in {\mathfrak K}_-$.

\noindent
{\bf Note}\,\,There is no diagonal matrix expressions for
$\delta_*(z{+}\frac{1}{i\h}u{\ctt}v)$, as 
$\int_{\mathbb R}e^{s(z{+}n{+}\frac{1}{2})}ds$ diverges.

\subsubsection{Further remarks for diagonal matrix calculations}

Diagonal matrix calculus is the calculus based on the 
diagonal matrix expressions of
$e_*^{(s{+}it)(z{+}\frac{1}{i\h}u{\ctt}v)}$ 
and the termwise differentiations and integrations. 
Let $I_{\ctt}(K)=(a,b)$ be the exchanging interval of generic ordered
expression $K$.   
What is important is not an expression parameter, but the type $E$, $\bar E$ or $D$ of matrix, 
Thus, this is the procedure of taking limit of expression parameter
$K$ so that $\lim I_{\ctt}(K)$ tends to the whole line or $\emptyset$, that is, 
${E(K)mat}$-expression (resp. ${D(K)mat}$-expression,
$\overline{E}(K)mat$ ) is the diagonal matrix expression by 
the virtual expression parameter $K$ such that $I_{\ctt}(K)=(\infty,\infty)$
(resp. $I_{\ctt}(K)=(-\infty,\infty)$, $I_{\ctt}(K)=(-\infty,-\infty$)).

As diagonal matrices, we see the following:
 
\begin{thm}\label{unifiedrep}
$e_*^{(s{+}it)\frac{1}{i\h}u{\ctt}v}$ is expressed by diagonal
matrices 
$$
\begin{aligned}
\begin{matrix}
\medskip
{:}e_*^{(s{+}it)(z{+}\frac{1}{i\h}u{\ctt}v)}{:}_{_{E(K)mat}}=&
\sum_{k=0}^{\infty}E_{k,k}(K)e^{(s{+}it)(z{+}k{+}\frac{1}{2})}, \\
\medskip
{:}e_*^{(s{+}it)(z{+}\frac{1}{i\h}u{\ctt}v)}{:}_{_{D(K)mat}}=&
\sum_{n=-\infty}^{\infty}\tilde{D}_{n}(K)e^{(s{+}it)(z{+}n)},\\
\medskip
{:}e_*^{(s{+}it)(z{-}\frac{1}{i\h}u{\ctt}v)}{:}_{_{\overline{E}(K)mat}}=&
\sum_{k=0}^{\infty}\overline{E}_{k,k}(K)e^{(s{+}it)(z{-}k{-}\frac{1}{2})},\\
\end{matrix}
\end{aligned}
$$
where ${E}_{k,k}(K)$, $\overline{E}_{k,k}(K)$ equal to the ones given
by \eqref{EE}, and $\tilde{D}_{n}(K)={D}_{n,n}(K)$ if 
$K\in {\mathfrak K}_0$. 
\end{thm}
 
By this we see in particular  
$$
{:}(z{+}\frac{1}{i\h}u{\ctt}v){:}_{_{D(K)mat}}=\sum_{n=-\infty}^{\infty}(z{+}n)D_{n,n}(K).
$$
On the other hand, we proved in \cite{ommy6} that in generic
$K$-expression 
$$
D_{K}^{-1}(z{+}\frac{1}{i\h}u{\ctt}v)=\sum_{n=-\infty}^{\infty}\frac{1}{z{+}n}D_{n,n}(K)
$$
converges in $H\!ol({\mathbb C}^2)$ as partial fractions, and this is
the inverse of ${:}(z{+}\frac{1}{i\h}u{\ctt}v){:}_{_{D(K)mat}}$ as
diagonal matrix calculations.

\begin{thm}\label{weaksurprise}
In generic ordered expression, 
$\sum_{n{=}0}^{\infty}E_{n,n}(K)$, $\sum_{n{=}0}^{\infty}\overline{E}_{n,n}(K)$
and $\sum_{n{=}-\infty}^{\infty}D_{n,n}(K)$ represent $1$, which will
be denoted by ${:}1{:}_{Emat}$, ${:}1{:}_{{\overline{E}}mat}$,
${:}1{:}_{Dmat}$ respectively

Hence every element of the Weyl algebra is expressed various ways
depending on the expression parameter.   
\end{thm}
\noindent
{\bf Note}\,\, A diagonal matrix may be viewed as a ``field'' on a
discrete set ${\mathbb Z}$ or $\mathbb N$. Componentwise/brock-wise
calculation will become a powerful tool to analyze peculiar nature of 
$*$-exponential functions and their integrals, 
just as particle physics jumps to the field theory to
analyze creation/annihilation procedure.

\section{Several $*$-special functions}  
As applications of calculus involving diagonal matrix elements, we 
define several $*$-special functions.  

Ordinary beta function $B(x,y)$ is defined for ${\rm{Re}}\,x>0$, ${\rm{Re}}\,y>0$ by   
$$
B(x,y)=\int_0^1t^{x{-}1}(1{-}t)^{y{-}1}dt.
$$
Replacing $t$ by $e^s$, we define in generic $K$-expression, $*$-beta functions by 
\begin{equation}
\label{eq:beta}
{:}B_*(\alpha{\pm}\frac{\tau}{i\h}u{\ctt}v,y){:}_{_K}=
\int_{-\infty}^0
{:}e_*^{s(\alpha{\pm}\frac{\tau}{i\h}u{\ctt}v)}{:}_{_K} 
(1{-}e^{s})^{y{-}1}ds,\,\,{\rm{Re}}\,\,\alpha>-\frac{|\tau|}{2},
 \,\,{\rm{Re}}\,y>0.
\end{equation}
These converge uniformly on every compact domain of $\alpha$ to give 
$H\!ol({\mathbb C}^2)$-valued holomorphic functions, but noting that 
$\int_{-\infty}^0{:}e_*^{s(\alpha{\pm}\frac{1}{i\h}u{\ctt}v)}{:}_{_K}ds$
give two different inverses of $\alpha{\pm}\frac{\tau}{i\h}u{\ctt}v$, we
have to set 
\begin{equation}\label{beta000}
B_*(\alpha{+}\frac{\tau}{i\h}u{\ctt}v,1)= 
(\alpha{+}\frac{\tau}{i\h}u{\ctt}v)_{*+}^{-1},\quad 
B_*(\alpha{-}\frac{\tau}{i\h}u{\ctt}v,1)= 
(\alpha{-}\frac{\tau}{i\h}u{\ctt}v)_{*-}^{-1}{=}
-({-}\alpha{+}\frac{\tau}{i\h}u{\ctt}v)_{*-}^{-1}.
\end{equation}

\subsection{Star-gamma functions}
We define in generic $K$-expression  as follows:
\begin{equation}
\label{eq:Gamm000}
{:}\varGamma_*(z{\pm}\frac{1}{i\h}u{\ctt}v){:}_{_{K}}=
\int_{-\infty}^{\infty}e^{-e^{s}}
{:}e_*^{s(z{\pm}\frac{1}{i\h}u{\ctt}v)}{:}_{_K}ds, \quad 
{\rm{Re}}\,\,z>-\frac{1}{2}
\end{equation}
If ${\rm{Re}}\,z>-\frac{1}{2}$,  
this integral converges in a generic ordered expression  
to give an element of $H\!ol({\mathbb C}^2)$. 
As the usual gamma function, integration by parts gives the identity 
\begin{equation}
 \label{eq:func}
\varGamma_*(z{+}1\pm\frac{1}{i\h}u{\ctt}v){=}
(z\pm\frac{1}{i\h}u{\ctt}v){*}
\varGamma_*(z\pm\frac{1}{i\h}u{\ctt}v).
\end{equation}
Note here that $e^{-e^s}e^{\e s}$, $s\in \mathbb R$ is a rapidly
decreasing function for every $\e>0$. Its inverse Fourier transform is 
$$
\int_{\mathbb R}e^{-e^t}e^{\e t}e^{it\xi}\dbar{t}=\frac{1}{\sqrt{2\pi}}\varGamma(i\xi{+}\e).
$$
Moreover, $e^{-e^s}$ is a tempered distribution. Hence its inverse Fourier transform is
$\lim_{\e\downarrow 0}\frac{1}{\sqrt{2\pi}}\varGamma(i\xi{+}\e)$.
Thus \eqref{tempdist} gives  
\begin{equation}\label{formgamma}
{:}\varGamma_*(z{\pm}\frac{1}{i\h}u{\ctt}v{}){:}_{_{K}}=
\frac{1}{2\pi}
\lim_{\e\downarrow 0}\int_{\mathbb R}
\varGamma(i\xi{+}\e){:}\delta_*(iz{\pm}\frac{1}{\h}u{\ctt}v{-}\xi){:}_{_K}d\xi
\end{equation}

\bigskip
As $\varGamma(z)$ extends on ${\mathbb C}{\setminus}\{-\mathbb N\}$, 
recalling \eqref{formaldist} one may consider   
\begin{equation}\label{algericgamma}
\varGamma_*^{C}(z{\pm}\frac{1}{i\h}u{\ctt}v)=
\frac{1}{2\pi i}\int_{C}\varGamma(iw)\delta_*(iz{\pm}\frac{1}{\h}u{\ctt}v{-}w)dw
\end{equation}
to analyze the interaction of singularities of $\varGamma(iw)$ and 
$\delta_*(iz{\pm}\frac{1}{\h}u{\ctt}v{-}w)$.

\subsubsection{Analytic continuation of 
${:}\varGamma_*(z{\pm}\frac{1}{i\h}u{\ctt}v){:}_{_K}$
in $H\!ol({\mathbb C}^2)$} 

As $\delta_*(z\pm\frac{1}{i\h}u{\ctt}v)$ is analytically continued,
the formula \eqref{formgamma} gives also the analytic continuation of 
$\varGamma_*(z{\pm}\frac{\tau}{i\h}u{\ctt}v)$, but as 
$*$-gamma function is defined by the integral form, it is easy
to obtain the formula for the analytic continuation.
\begin{prop}
 \label{conanal}
$\varGamma_*(z\pm\frac{1}{i\h}u{\ctt}v)$ 
extends to an $H\!ol({\mathbb C}^2)$-valued holomorphic function on 
$$
 z\in {\mathbb C}{\setminus}\{-({\mathbb N}{+}\frac{1}{2})\}
$$ 
with simple poles.  
\end{prop}

\noindent
{\bf Proof}\,\,Split the integral 
into two parts $\int_{-\infty}^0 +\int_0^{\infty}$.  It is easy 
to see that 
$\int_0^{\infty}
e^{-e^t}e_*^{t(z{\pm}\frac{1}{i\h}u{\ctt}v)}dt$ is entire. 
 Thus we have only to show for the first term. 
For 
every $z$, choose an positive integer $n$ such that 
${\rm{Re}}(z+n)>0$. Set 
$$
e^{-e^{t}}= 1{-}e^t+\frac{1}{2}e^{2t}+\cdots+
            \frac{(-1)^n}{n!}e^{nt}+e^{nt}F_n(t),\quad |F_n(t)|\leq C.
$$
Then, it is easy to see 
$R_n(z{\pm}\frac{1}{i\h}u{\ctt}v)=
\int_{-\infty}^0 
F_n(t)e^{nt}e_*^{t(z{\pm}\frac{1}{i\h}u{\ctt}v)}dt$ 
converges, and hence   
$$
\begin{aligned}
\int_{-\infty}^0
&e^{-e^t}e_*^{t(z{\pm}\frac{1}{i\h}u{\ctt}v)}dt\\
&=(z{\pm}\frac{1}{i\h}u{\ctt}v)_{*\pm}^{-1}-
(1{+}z{\pm}\frac{1}{i\h}u{\ctt}v)_{*\pm}^{-1}+\cdots
+\frac{(-1)^n}{n!}
(n{+}z{\pm}\frac{1}{i\h}u{\ctt}v)_{*\pm}^{-1}+
R_n(z{\pm}\frac{1}{i\h}u{\ctt}v). 
\end{aligned}
$$
The result follows immediately.\hfill $\Box$

\subsubsection{Residues of 
$\varGamma_*(z\pm\frac{1}{i\h}u{\ctt}v)$}

The proof of Proposition\,\,\ref{conanal} gives also the 
formula of residues. 
By Theorem\,\ref{deltacont}, we have  
$$
{\rm{Res}}\big({:}\varGamma_*(z{+}\frac{1}{i\h}u{\ctt}v){:}_{_K}, -(n+\frac{1}{2})\big)
{=}\sum_{k=0}^n(-1)^k\frac{1}{(n{-}k)!}E_{k,k}(K)
$$
$$
{\rm{Res}}\big({:}\varGamma_*(z{-}\frac{1}{i\h}u{\ctt}v){:}_{_K}, -(n+\frac{1}{2})\big)
{=}\sum_{k=0}^n(-1)^k\frac{1}{(n{-}k)!}\overline{E}_{k,k}(K).
$$

Using this we have 
\begin{prop}
In generic $K$-ordered expression, 
$$
\frac{1}{\varGamma(z{+}\frac{1}{2})}\varGamma_*(z{+}\frac{1}{i\h}u{\ctt}v),\quad 
\frac{1}{\varGamma(z{+}\frac{1}{2})}\varGamma_*(z{-}\frac{1}{i\h}u{\ctt}v)
$$

are  $H\!ol({\mathbb C}^2)$-valued entire functions of $z$.  
\end{prop} 
Note that all zero's of
$\frac{1}{\varGamma(z{+}\frac{1}{2})}$ are 
eliminated by the singularities of $\varGamma(z{\pm}\frac{1}{i\h}u{\ctt}v)$. 
Hence 
$$
\frac{1}{\varGamma(z{+}\frac{1}{2})}\varGamma_*(z{\pm}\frac{1}{i\h}u{\ctt}v)
$$ 
has no vanishing point. This suggests that the $*$-inverse of  
$\varGamma_*(z{\pm}\frac{1}{i\h}u{\ctt}v)$ may be treated in this way.

\subsection{Diagonal matrix expressions of 
$\varGamma_*(z{+}\frac{1}{i\h}u{\ctt}v)$}
Let 
$I_{\ctt}(K)=(a,b)$ is the exchanging interval of ${:}e_*^{z\frac{1}{i\h}u{\ctt}v}{:}_{_K}$.
Applying \eqref{Fourier} for $t=0$ in a generic $K$-expression, we have
the hybrid matrix expression   
$$
\begin{aligned}
&{:}\varGamma_*(z{+}\frac{1}{i\h}u{\ctt}v){:}_{_K}=
\int_{-\infty}^{\infty}e^{-e^{s}}
{:}e_*^{s(z{+}\frac{1}{i\h}u{\ctt}v)}{:}_{_{K}}ds\\
&{=}\!\int_{-\infty}^{a}\!\!\!e^{-e^{s}}\!\sum_{n=0}^{\infty}e^{s(z{+}(n{+}\frac{1}{2}))}E_{n,n}(K)ds{+}
\!\int_{a}^{b}\!\!e^{-e^s}\!\!\sum_{n=-\infty}^{\infty}\!e^{s(z{+}n)}\tilde{D}_{n}(K)ds{+}
\!\int_{b}^{\infty}\!\!e^{-e^{s}}\!\sum_{n=0}^{\infty}
e^{s(z{-}(n{+}\frac{1}{2}))}\overline{E}_{n,n}(K)ds.
\end{aligned}
$$
Using this, one may take diagonal matrix expressions, but here  
we give first a justification of the diagonal matrix calculations.

\bigskip
Note that for every $\sigma$
$$
\begin{aligned}
&\int_{-\infty}^{\infty}e^{-e^{s}}
{:}e_*^{s(z{+}\frac{1}{i\h}u{\ctt}v)}{:}_{_K}ds\\
&=
\int_{-\infty}^{a{+}\sigma}\!\!e^{-e^{s}}
{:}e_*^{s(z{+}\frac{1}{i\h}u{\ctt}v)}{:}_{_K}ds
{+}\int_{a{+}\sigma}^{b{+}\sigma}\!\!e^{-e^{s}}
{:}e_*^{s(z{+}\frac{1}{i\h}u{\ctt}v)}{:}_{_K}ds
{+}
\int_{b{+}\sigma}^{\infty}\!\!e^{-e^{s}}
{:}e_*^{s(z{+}\frac{1}{i\h}u{\ctt}v)}{:}_{_K}ds\\
&=
\int_{-\infty}^{a}\!\!e^{-e^{s{+}\sigma}}
{:}e_*^{(s{+}\sigma)(z{+}\frac{1}{i\h}u{\ctt}v)}{:}_{_K}ds
{+}\int_{a}^{b}\!\!e^{-e^{s{+}\sigma}}
{:}e_*^{(s{+}\sigma)(z{+}\frac{1}{i\h}u{\ctt}v)}{:}_{_K}ds
{+}
\int_{b}^{\infty}\!\!e^{-e^{s{+}\sigma}}
{:}e_*^{(s{+}\sigma)(z{+}\frac{1}{i\h}u{\ctt}v)}{:}_{_K}ds
\end{aligned}
$$
As the $*$-product
$e_*^{\sigma(z{+}\frac{1}{i\h}u{\ctt}v)}{*}H_*$ is defined by 
the analytic solution of the evolution equation, we have  
$$
\begin{aligned}
&\int_{-\infty}^{a}e^{-e^{s{+}\sigma}}
{:}e_*^{(s{+}\sigma)(z{+}\frac{1}{i\h}u{\ctt}v)}{:}_{_K}ds=
{:}e_*^{\sigma(z{+}\frac{1}{i\h}u{\ctt}v)}{:}_{_K}
{*}\int_{-\infty}^{a}e^{-e^{s{+}\sigma}}
{:}e_*^{s(z{+}\frac{1}{i\h}u{\ctt}v)}{:}_{_K}ds\\
&\int_{a}^{b}\!\!e^{-e^{s{+}\sigma}}
{:}e_*^{(s{+}\sigma)(z{+}\frac{1}{i\h}u{\ctt}v)}{:}_{_K}ds
={:}e_*^{\sigma(z{+}\frac{1}{i\h}u{\ctt}v)}
{*}\int_{a}^{b}\!\!e^{-e^{s{+}\sigma}}e_*^{s(z{+}\frac{1}{i\h}u{\ctt}v)}{:}_{_K}ds\\
&
\int_{b}^{\infty}\!\!e^{-e^{s{+}\sigma}}
{:}e_*^{(s{+}\sigma)(z{+}\frac{1}{i\h}u{\ctt}v)}{:}_{_K}ds
=
{:}e_*^{\sigma(z{+}\frac{1}{i\h}u{\ctt}v)}
{*}\int_{b}^{\infty}\!\!e^{-e^{s{+}\sigma}}
e_*^{s(z{+}\frac{1}{i\h}u{\ctt}v)}{:}_{_K}ds
\end{aligned}
$$
We see 
$$
\lim_{\sigma\to\infty}\int_{a}^{b}\!\!e^{-e^{s{+}\sigma}}
{:}e_*^{(s{+}\sigma)(z{+}\frac{1}{i\h}u{\ctt}v)}{:}_{_K}ds{=}0,\quad 
\lim_{\sigma\to\infty}\int_{b}^{\infty}\!\!e^{-e^{s{+}\sigma}}
{:}e_*^{(s{+}\sigma)(z{+}\frac{1}{i\h}u{\ctt}v)}{:}_{_K}ds=0
$$

\medskip
One of the remarkable feature of $*$-gamma function is that
$E(K)mat$-expression, and $\overline{E}(K)mat$-expression of   
$\varGamma_*(z{\pm}\frac{\tau}{i\h}u{\ctt}v)$ are  
obtained for generic $K$-expression. 
First, to obtain $E(K)mat$-expression, 
note that for ${\rm{Re}}\,z>-\frac{1}{2}$
$$
\lim_{\sigma\to\infty}\int_{-\infty}^{a{-}1{+}\sigma}e^{-e^{s}}
{:}e_*^{s(z{+}\frac{1}{i\h}u{\ctt}v)}{:}_{_K}ds
=\int_{-\infty}^{\infty}e^{-e^{s}}
{:}e_*^{s(z{+}\frac{1}{i\h}u{\ctt}v)}{:}_{_K}ds.
$$

We see 
$$
{:}e_*^{s(z{+}\frac{1}{i\h}u{\ctt}v)}{:}_{_K}=
\sum_{n=0}^{\infty}e^{s(z{+}(n{+}\frac{1}{2}))}E_{n,n}(K),\quad 
s<a{-}1.
$$
As this is the convergence of Taylor series at $w=0$, $w=e^t$, we see 
\begin{lem}\label{unifconv}
The convergence is uniform on $s<a{-}1$. 
\end{lem}
Termwise integration gives 
$$
\int_{-\infty}^{a{-}1}e^{-e^{s{+}\sigma}}
{:}e_*^{s(z{+}\frac{1}{i\h}u{\ctt}v)}{:}_{_K}ds=
\sum_{n=0}^{\infty}
\int_{-\infty}^{a{-}1}e^{-e^{s{+}\sigma}}
e^{s(z{+}(n{+}\frac{1}{2}))}ds E_{n,n}(K).
$$
in $H\!ol({\mathbb C}^2)$, but the next one holds only in matrix
$$
{:}e_*^{\sigma(z{+}\frac{1}{i\h}u{\ctt}v)}{:}_{_K}
{*}\int_{-\infty}^{a{-}1}e^{-e^{s{+}\sigma}}
{:}e_*^{s(z{+}\frac{1}{i\h}u{\ctt}v)}{:}_{_K}ds=
\sum_{n=0}^{\infty}
\int_{-\infty}^{a{-}1}e^{-e^{s{+}\sigma}}
e^{s(z{+}(n{+}\frac{1}{2}))}ds 
{:}e_*^{\sigma(z{+}\frac{1}{i\h}u{\ctt}v)}{:}_{_K}{*}E_{n,n}(K).
$$
It is easy to see 
${:}e_*^{\sigma(z{+}\frac{1}{i\h}u{\ctt}v)}{:}_{_K}{*}E_{n,n}(K)=
e^{\sigma(z{+}\frac{1}{2})}E_{n,n}(K)$.

By using the Lebesgue dominated convergence theorem, we see 
$$
\int_{-\infty}^{\infty}e^{-e^{s}}
{:}e_*^{s(z{+}\frac{1}{i\h}u{\ctt}v)}{:}_{_K}ds=
\sum_{n=0}^{\infty}\int_{-\infty}^{\infty}e^{-e^{s}}
e^{s(z{+}(n{+}\frac{1}{2}))}ds E_{n,n}(K)
$$

Hence, we have in generic $K$-expression,
$$
{:}\varGamma_*(z{+}\frac{1}{i\h}u{\ctt}v){:}_{_K}=
\sum_{n=0}^{\infty}\varGamma(z{+}(n{+}\frac{1}{2}))E_{n,n}(K)
$$
but the r.h.s. converges only in diagonal matrices. Hence we denote
this by 
\begin{equation}\label{matgamma}
{:}\varGamma_*(z{+}\frac{1}{i\h}u{\ctt}v){:}_{_{E(K)mat}}=
\sum_{n=0}^{\infty}\varGamma(z{+}(n{+}\frac{1}{2}))E_{n,n}(K)=
\varGamma(z{+}\frac{1}{2})\sum_{n=0}^{\infty}(z{+}\frac{1}{2})_{n}E_{n,n}(K),
\end{equation}
where $(a)_n=a(a{+}1)\cdots(a{+}n{-}1)$, $(a)_0=1$.

Recall that $E(K)mat$-expression is the expression by the virtual
expression parameter such that $I_{\ctt}(K)=(\infty,\infty)$.  

\bigskip
$\overline{E}(K)mat$-expression is obtained by changing the orientation
of the integration. For ${\rm{Re}}\,z>-\frac{1}{2}$ 
$$
\lim_{\sigma\to\infty}\int_{b+1-\sigma}^{\infty}e^{-e^{s}}
{:}e_*^{s(z{+}\frac{1}{i\h}u{\ctt}v)}{:}_{_K}ds
=\int_{-\infty}^{\infty}e^{-e^{s}}
{:}e_*^{s(z{+}\frac{1}{i\h}u{\ctt}v)}{:}_{_K}ds.
$$
$$
\begin{aligned}
\int_{b+1-\sigma}^{\infty }e^{-e^{s}}
&{:}e_*^{s(z{+}\frac{1}{i\h}u{\ctt}v)}{:}_{_K}ds=
\int_{b{+}1}^{\infty}
e^{-e^{s-\sigma}}
{:}e_*^{(s{-}\sigma)(z{+}\frac{1}{i\h}u{\ctt}v)}{:}_{_K}ds\\
&=
{:}e_*^{-\sigma(z{+}\frac{1}{i\h}u{\ctt}v)}{:}_{_K}
{*}\int_{b{+}1}^{\infty}e^{-e^{s{-}\sigma}}
{:}e_*^{s(z{+}\frac{1}{i\h}u{\ctt}v)}{:}_{_K}ds\\
&=
{:}e_*^{-\sigma(z{+}\frac{1}{i\h}u{\ctt}v)}{:}_{_K}
{*}\int_{b{+}1}^{\infty}e^{-e^{s{-}\sigma}}
\sum_{n=0}^{\infty}e^{s(z{-}n{-}\frac{1}{2})}ds\overline{E}_{n,n}(K)\\
&=
\sum_{n=0}^{\infty}\int_{b{+}1-\sigma}^{\infty}e^{-e^{s}}
e^{s(z{-}n{-}\frac{1}{2})}ds\overline{E}_{n,n}(K).
\end{aligned}
$$
The last equality holds as diagonal matrices. Hence by the
componentwise convergence we have 
$$
{:}\varGamma_*(z{+}\frac{1}{i\h}u{\ctt}v){:}_{_{\overline{E}(K)mat}}
=
\sum_{n=0}^{\infty}\int_{-\infty}^{\infty}e^{-e^{s}}
e^{s(z{-}n{-}\frac{1}{2})}ds\overline{E}_{n,n}(K)=
\sum_{n=0}^{\infty}\varGamma(z{-}n{-}\frac{1}{2})\overline{E}_{n,n}(K).
$$

Since 
$\varGamma(z{-}n{-}\frac{1}{2}){=}(z{+}\frac{1}{2})_{-n}^{-1}\varGamma(z{+}\frac{1}{2})$
where $(a)_{-n}{=}(a{-}1)(a{-}2)\cdots(a{-}n)$, 
we have in generic $K$-expression,
\begin{equation}\label{matbargamma}
{:}\varGamma_*(z{+}\frac{1}{i\h}u{\ctt}v){:}_{_K}=
\sum_{n=0}^{\infty}\varGamma(z{-}(n{+}\frac{1}{2}))\overline{E}_{n,n}(K)=
\varGamma(z{+}\frac{1}{2})\sum_{n=0}^{\infty}(z{+}\frac{1}{2})^{-1}_{-n}\overline{E}_{n,n}(K).
\end{equation}

Recall again that $\overline{E}(K)mat$-expression is the expression by the virtual
expression parameter such that $I_{\ctt}(K)=(-\infty,-\infty)$.  

\bigskip
Similarly, ${D}(K)mat$-expression 
${:}\varGamma_*(z{+}\frac{1}{i\h}u{\ctt}v){:}_{_{D(K)mat}}$ is 
\begin{equation}\label{Dmat}
\begin{aligned}
{:}\varGamma_*(z{+}\frac{1}{i\h}u{\ctt}v){:}_{_{D(K)mat}}&=
\lim_{(a,b)\to(-\infty,\infty)}\int_{a}^{b}e^{-e^s}\sum_ne^{s(z{+}n)}ds\\
&=
\sum_{n}\varGamma(z{+}n)D_{n,n}(K)=\varGamma(z)\sum_{n=-\infty}^{\infty}(z)_n^{sgn(n)1}D_{n,n}(K).
\end{aligned}
\end{equation}
Hence this can not converge in $H\!ol({\mathbb C}^2)$, as $(z)_{n+1}\sim n!$ for $n>0$.

\bigskip
Computations for 
${:}\varGamma_*(z{-}\frac{1}{i\h}u{\ctt}v){:}_{_{E(K)mat}}$ is 
 parallel to the above, but a little care is required, since the direction of $s$, which
may be viewed as the time, is reversed. We summarize the results
below:
\begin{equation}\label{totresult}
\begin{aligned}
{:}\varGamma_*(z{+}\frac{1}{i\h}u{\ctt}v){:}_{_{{E}(K)mat}}
&{=}\sum_{n=0}^{\infty}\varGamma(z{+}n{+}\frac{1}{2}){E}_{n,n}(K)
 {=}\varGamma(z{+}\frac{1}{2})\sum_{n=0}^{\infty}(z{+}\frac{1}{2})_n{E}_{n,n}(K)\\
{:}\varGamma_*(z{+}\frac{1}{i\h}u{\ctt}v){:}_{_{\overline{E}(K)mat}}
&{=}\sum_{n=0}^{\infty}\varGamma(z{-}n{-}\frac{1}{2})\overline{E}_{n,n}(K)
 {=}\varGamma(z{+}\frac{1}{2})\sum_{n=0}^{\infty}(z{+}\frac{1}{2})_{-n}^{-1}\overline{E}_{n,n}(K)\\
{:}\varGamma_*(z{-}\frac{1}{i\h}u{\ctt}v){:}_{_{{E}(K)mat}}
&{=}\sum_{n=0}^{\infty}\varGamma(z{-}n{-}\frac{1}{2}){E}_{n,n}(K)
 {=}\varGamma(z{+}\frac{1}{2})\sum_{n=0}^{\infty}(z{+}\frac{1}{2})_{-n}^{-1}{E}_{n,n}(K)\\
{:}\varGamma_*(z{-}\frac{1}{i\h}u{\ctt}v){:}_{_{\overline{E}(K)mat}}
&{=}\sum_{n=0}^{\infty}\varGamma(z{+}n{+}\frac{1}{2})\overline{E}_{n,n}(K)
 {=}\varGamma(z{+}\frac{1}{2})\sum_{n=0}^{\infty}(z{+}\frac{1}{2})_n\overline{E}_{n,n}(K).\\
\end{aligned}
\end{equation}

\subsubsection{The inverse $\varGamma_*(z{\pm}\frac{1}{i\h}u{\ctt}v)^{-1}$}

Diagonal matrix calculation is useful to find the formula of inverses.
It is easy to see that the inverse of  
$\sum_n(z{+}\frac{1}{2})_{n}E_{n,n}(K)$ in diagonal matrices is
given by $\sum_n(z{+}\frac{1}{2})_{n}^{-1}E_{n,n}(K)$.  

By Theorem\,\ref{Nicecoeff} we see the next
\begin{thm}\label{invGamma}
If $K{\in}{\mathfrak K}_+$, then  
$$
{:}\varGamma_*^{-1}(z{+}\frac{1}{i\h}u{\ctt}v){:}_{_{E(K)mat}}
=\varGamma(z{+}\frac{1}{2})^{-1}\sum_n(z{+}\frac{1}{2})_{n}^{-1}E_{n,n}(K),
$$
converge in $H\!ol({\mathbb C}^2)$, and every component 
$\varGamma(z{+}\frac{1}{2})^{-1}(z{+}\frac{1}{2})_{n}^{-1}$ is an
entire function. 
Similarly, if 
$K{\in}{\mathfrak K}_-$, then
$$
{:}\varGamma_*^{-1}(z{-}\frac{1}{i\h}u{\ctt}v){:}_{_{\overline{E}(K)mat}}
=\varGamma(z{+}\frac{1}{2})^{-1}\sum_n(z{+}\frac{1}{2})_{n}^{-1}\overline{E}_{n,n}(K)
$$
converges in $H\!ol({\mathbb C}^2)$ and every component 
$\varGamma(z{+}\frac{1}{2})^{-1}(z{+}\frac{1}{2})_{n}^{-1}$ is an
entire function.
\end{thm}

\noindent
{\bf Proof}\,\,\,Fix $z$ so that $(z{+}\frac{1}{2})_{n}\not=0$. 
Since 
$\sum_n(z{\pm}\frac{1}{2})_{n}^{-1}e^{in\theta}$ is viewed as the Fourier
series of a smooth function $f(\theta)$ on $S^1$, one can apply
Theorem\,\ref{Nicecoeff}.\hfill $\Box$  

As the byproduct we have 
\begin{cor}\label{inversegamma}
If $K{\in}{\mathfrak K}_{+}$, $($ {\rm{resp.  }}  ${\mathfrak K}_{-}$ $)$, 
then
$$
\varGamma(z{+}\frac{1}{2})^{-1}\sum_n(z{+}\frac{1}{2})_{n}^{-1}{E}_{n,n}(K),\quad 
{\rm{resp. }}  
\quad \varGamma(z{+}\frac{1}{2})^{-1}\sum_n(z{+}\frac{1}{2})_{n}^{-1}\overline{E}_{n,n}(K)
$$  
is a genuine inverse of 
$\varGamma_*(z{+}\frac{1}{i\h}u{\ctt}v)$ 
$($ {\rm{resp.}  } $\varGamma_*(z{-}\frac{1}{i\h}u{\ctt}v)$ $)$.
\end{cor}

Note that elements such as 
$$
\varGamma(z{+}\frac{1}{2})^{-1}\sum_n(z{+}\frac{1}{2})_{n}^{-1}{E}_{n,n}(K){+}
\varGamma(z{+}\frac{1}{2})^{-1}\sum_n(z{+}\frac{1}{2})_{n}^{-1}\overline{E}_{n,n}(K)
$$
makes sense as diagonal matrices, which will be called a {\bf hybrid}
matrix expression.  It is clear that 
$$
\sum_{n=0}^{\infty}\!\varGamma(z{+}n{+}\frac{1}{2})({E}_{n,n}(K){+}\overline{E}_{n,n}(K)){*}
\sum_{n=0}^{\infty}\!\varGamma(z{+}n{+}\frac{1}{2})^{-1}({E}_{n,n}(K){+}\overline{E}_{n,n}(K))
{=}
\sum_{n=0}^{\infty}\!({E}_{n,n}(K){+}\overline{E}_{n,n}(K))
$$

\bigskip
In the next section, we discuss the $*$-inverse of 
$\varGamma_*(z{\pm}\frac{1}{i\h}u{\ctt}v)$ by using the infinite
product formula.

\section{The infinite product formula of 
$\varGamma_*(z{\pm}\frac{1}{i\h}u{\ctt}v)$} 
Recall 
$$
{:}\varGamma_*(z{+}\frac{1}{i\h}u{\ctt}v){:}_{_{E(K)mat}}{=}
\sum_{k=0}^{\infty}\varGamma(z{+}k{+}\frac{1}{2})E_{k,k}(K).
$$
The infinite product formula of ordinary gamma function gives 
$$
{:}\varGamma_*(z{+}\frac{1}{i\h}u{\ctt}v){:}_{_{E(K)mat}}{=}
\sum_{k=0}^{\infty}\Big(e^{-\gamma(z{+}k{+}\frac{1}{2})}(z{+}k{+}\frac{1}{2})^{-1}
\prod_{\ell=0}^{\infty}
\big(1{+}\frac{1}{\ell}(z{+}k{+}\frac{1}{2})^{-1}e^{\frac{1}{\ell}(z{+}k{+}\frac{1}{2})}\big)
\Big)E_{k,k}(K).
$$
It is a remarkable feature of diagonal matrix calculation that the  
multiplications commute with the summations allows:
$$
\begin{aligned}
{:}\varGamma_*(z{+}\frac{1}{i\h}u{\ctt}v){:}_{_{E(K)mat}}&{=}
\prod_{\ell=0}^{\infty}\Big(\sum_{k=0}^{\infty}e^{-\gamma(z{+}k{+}\frac{1}{2})}
(z{+}k{+}\frac{1}{2})^{-1}
\big(1{+}\frac{1}{\ell}(z{+}k{+}\frac{1}{2})\big)^{-1}e^{\frac{1}{\ell}(z{+}k{+}\frac{1}{2})}
E_{k,k}(K)\Big)\\
&{=}
\prod_{\ell=0}^{\infty}
\Big({:}e^{-\gamma(z{+}\frac{1}{i\h}u{\ctt}v)}{*}
(z{+}\frac{1}{i\h}u{\ctt}v)_{*+}^{-1}{*}
\big(1{+}\frac{1}{\ell}(z{+}\frac{1}{i\h}u{\ctt}v)\big)_{*+}^{-1}
e_*^{\frac{1}{\ell}(z{+}\frac{1}{i\h}u{\ctt}v)}{:}_{_{Kmat}}\Big).
\end{aligned}
$$

\bigskip
The next Lemma is crucial to obtain the infinite product formula in
$H\!ol({\mathbb C}^2)$. 
\begin{lem}
  \label{prod10}
$B_*(z\pm\frac{1}{i\h}u{\ctt}v, n{+}1){=}
n!\prod_{k=0}^n{*}(k{+}z\pm\frac{1}{i\h}u{\ctt}v)^{-1}_{*\pm}$,\,\,
${\rm{Re}}\,z{>}-\frac{1}{2}$.
\end{lem}

\noindent
{\bf Proof}\,\,\,\,The r.h.s. of the above equality will be denoted by 
$\frac{n!}{\{z\pm\frac{1}{i\h}u{\ctt}v\}^{(\pm)}_{*{n{+}1}}}$.

The case $n=0$ is given by \eqref{beta000}. Suppose this is
true for $n$. For the case $n{+}1$, the definition \eqref{eq:Gamm000} and the
induction assumption give   
$$
\begin{aligned}
B_*(z\pm\frac{1}{\h i}u{\ctt}v, n{+}2)
=&\int_{-\infty}^{0}
e_*^{\tau(z\pm\frac{1}{i\h}u{\ctt}v)}
(1{-}e^{\tau})(1{-}e^{\tau})^nd\tau \\
=&\frac{n!}{\{z\pm\frac{1}{\h i}u{\ctt}v\}^{(\pm)}_{*{n{+}1}}}
-\frac{n!}{\{1{+}z\pm\frac{1}{\h i}u{\ctt}v\}^{(\pm)}_{*{n{+}1}}}. 
\end{aligned}
$$
By noting the standard resolvent identity 
$$
\frac{n!}{A{*}(A{+}1){*}\cdots{*}(A{+}n)}-
\frac{n!}{(A{+}1)(A{+}2){*}\cdots{*}(A{+}n{+}1)}=\frac{(n{+}1)!}{A{*}(A{+}1){*}\cdots{*}(A{+}n{+}1)}
$$
we have  that 
$$
B_*(z\pm\frac{1}{\h i}u{\ctt}v, n{+}2)=
\frac{(n{+}1)!}{\{z\pm\frac{1}{\h i}u{\ctt}v\}^{(\pm)}_{*{n{+}2}}}.
$$
Replace $n{+}2$ by $n{+}1$, we get the formula.
${}$ \hfill $\Box$

\bigskip
Since ${*}$-products of inverses are given by summations, we have also
the following: 
\begin{prop}\label{invconti}
$\frac{n!}{\{z\pm\frac{1}{\h i}u{\ctt}v\}^{(\pm)}_{*{n{+}1}}}$ is
analytically continued on $\mathbb C{\setminus}\{-({\mathbb N}{+}\frac{1}{2})\}$.
\end{prop}

\bigskip
The infinite product formula for 
$*$-gamma function will be given by this formula. By Lemma\,\ref{prod10}, we see 
$$
\int_{-\infty}^0
e_*^{\tau(z\pm\frac{1}{\h i}u{\ctt}v)}(1-e^{\tau})^nd\tau =
\frac{n!}{\{z\pm\frac{1}{\h i}u{\ctt}v\}^{(\pm)}_{*{n{+}1}}},
\quad {\rm{Re}}\,z>-\frac{1}{2}.
$$
Replacing $e^{\tau}$ by $\frac{1}{n}e^{\tau'}$, namely setting 
$\tau=\tau'-\log n$ in the left hand side and 
multiplying $e_*^{(\log n)(z\pm\frac{1}{\h i}u{\ctt}v)}$ to the both sides, 
we have in generic ordered expression 
\begin{equation}
 \label{eq:form01}
\int_{-\infty}^{\log n}e_*^{\tau'(z\pm\frac{1}{i\h}u{\ctt}v)}
(1{-}\frac{1}{n}e^{\tau'})^nd\tau' 
=\frac{n!}{\{z\pm\frac{1}{\h}u{\ctt}v\}^{(\pm)}_{*{n{+}1}}}
{*}e_*^{(\log n)(z\pm\frac{1}{\h}u{\ctt}v)}.  
\end{equation}
\begin{thm}
\label{keytop}  
In generic ordered expression, the left hand side converges 
 with $n{\to}\infty$ uniformly on each compact set of $z$ to 
$$\int_{-\infty}^{\infty}
e_*^{\tau'(z\pm\frac{1}{i\h}u{\ctt}v)}e^{-e^{\tau'}}d\tau'
=\varGamma_*(z\pm\frac{1}{i\h}u{\ctt}v)  
$$ 
in the space $H{\!o}l({\mathbb C}^2)$.  In fact, the convergence is
uniform on each domain $-\frac{1}{2}<{\rm{Re}}\,\,z<C$.
\end{thm}

\noindent
{\bf Proof}\,\,\, It is easy to see that 
$$
\lim_{n\to\infty}\int_{-\infty}^{0}e_*^{\tau'(z\pm\frac{1}{i\h}u{\ctt}v)}
(1{-}\frac{1}{n}e^{\tau'})^nd\tau'= 
\int_{-\infty}^{0}e_*^{\tau'(z\pm\frac{1}{i\h}u{\ctt}v)}e^{-e^{\tau'}}d\tau',
\quad {\rm{Re}}\,\,z>-\frac{1}{2}.
$$
Hence we have only to show for the integral $\int_{0}^{\log n}d\tau'$. 
For every positive integer $\ell$, consider the 
nonnegative function $x^{\ell}(1{-}\frac{x}{n})^n$ on
the interval $[0,n]$ for $n>>0$. At $x=\frac{n\ell}{n{+}\ell}$ 
this attains the 
maximum  $(\frac{n\ell}{n{+}\ell})^{\ell}(1{-}\frac{\ell}{n{+}\ell})^n$.
Since  
$$
\lim_{n\to\infty}(\frac{n\ell}{n{+}\ell})^{\ell}(1{-}\frac{\ell}{n{+}\ell})^n
=\ell^{\ell}(e^{-\ell})^{-\ell}=\ell^{\ell}e^{\ell^2},
$$
$x^{\ell}(1{-}\frac{x}{n})^n$ on the interval $[0,n]$ is dominated
uniformly by $C'\ell^{\ell}e^{\ell^2}$, ($C'>1$). Hence one may assume 
$$
(1{-}\frac{e^{\tau'}}{n})^n<C'\ell^{\ell}e^{\ell^2}e^{-\ell\tau'},
\quad n\gg 0.
$$

By using the Lebesgue dominated convergence theorem, we see 
that for ${\rm{Re}}z<\ell{+}\frac{1}{2}$, 
$$
\lim_{n\to\infty}
\int_{0}^{\log n}
e_*^{\tau'(z\pm\frac{1}{\h i}u{\ctt}v)}(1{-}\frac{e^{\tau'}}{n})^nd\tau'
{=}
\int_{0}^{\infty}
e_*^{\tau'(z\pm\frac{1}{\h i}u{\ctt}v)}e^{-e^{\tau'}}d\tau'
$$
in the space $H{\!o}l({\mathbb C}^2)$.
\hfill $\Box$

\medskip
Note that the proof can be applied to the case stated below:
\begin{prop}\label{loose}
Let $\{a_n\}_n$ be a converging series such that $\lim_n a_n{=}a$. 
$$
\begin{aligned}
\lim_{n\to\infty}
e_*^{a_n(z\pm\frac{1}{\h i}u{\ctt}v)}{*}\int_{-\infty}^{\log n}
e_*^{\tau'(z\pm\frac{1}{\h i}u{\ctt}v)}e^{-e^{\tau'}}d\tau'
&{=}
\int_{-\infty}^{\infty}
e_*^{a(z\pm\frac{1}{\h i}u{\ctt}v)}{*}
e_*^{\tau'(z\pm\frac{1}{\h i}u{\ctt}v)}e^{-e^{\tau'}}d\tau'\\
&=e_*^{a(z\pm\frac{1}{\h i}u{\ctt}v)}{*}
\int_{-\infty}^{\infty}
e_*^{\tau'(z\pm\frac{1}{\h i}u{\ctt}v)}e^{-e^{\tau'}}d\tau', \quad {\rm{Re}}\,\,z>-\frac{1}{2}.
\end{aligned}
$$
\end{prop}

\noindent
{\bf Proof}\,\,\,For the second equality, note that 
$$
\begin{aligned}
\frac{d}{da}\int_{-\infty}^{\infty}
e_*^{a(z\pm\frac{1}{\h i}u{\ctt}v)}{*}
e_*^{\tau'(z\pm\frac{1}{\h i}u{\ctt}v)}e^{-e^{\tau'}}d\tau'&=
\int_{-\infty}^{\infty}
(z\pm\frac{1}{\h i}u{\ctt}v){*}
e_*^{a(z\pm\frac{1}{\h i}u{\ctt}v)}{*}
e_*^{\tau'(z\pm\frac{1}{\h i}u{\ctt}v)}e^{-e^{\tau'}}d\tau'\\
&=
(z\pm\frac{1}{\h i}u{\ctt}v){*}
\int_{-\infty}^{\infty}e_*^{a(z\pm\frac{1}{\h i}u{\ctt}v)}{*}\!
e_*^{\tau'(z\pm\frac{1}{\h i}u{\ctt}v)}e^{-e^{\tau'}}d\tau'.
\end{aligned}
$$
As 
$e_*^{a(z\pm\frac{1}{\h i}u{\ctt}v)}{*}
e_*^{\tau'(z\pm\frac{1}{\h i}u{\ctt}v)}e^{-e^{\tau'}}$ is 
real analytic w.r.t. $a$ and the integral converges, we see that 
$$
\int_{-\infty}^{\infty}
(z\pm\frac{1}{\h i}u{\ctt}v){*}
e_*^{a(z\pm\frac{1}{\h i}u{\ctt}v)}{*}
e_*^{\tau'(z\pm\frac{1}{\h i}u{\ctt}v)}e^{-e^{\tau'}}d\tau'
$$ 
is a real analytic solution of the evolution equation 
$\frac{d}{da}f_a=(z\pm\frac{1}{\h i}u{\ctt}v){*}f_a$ with 
the initial data 
$f_0=\int_{-\infty}^{\infty}
e_*^{\tau'(z\pm\frac{1}{\h i}u{\ctt}v)}e^{-e^{\tau'}}d\tau'$.
Hence we have the second equality.\hfill $\Box$

\bigskip
Since 
$$
\frac{n!}{\{z\pm\frac{1}{\h i}u{\ctt}v\}^{(\pm)}_{*{n{+}1}}}=
(z\pm\frac{1}{\h i}u{\ctt}v)_{*\pm}^{-1}
{*}\prod_{k=1}^n{*}
(1{+}\frac{1}{k}(z\pm\frac{1}{\h i}u{\ctt}v))_{*\pm}^{-1},
$$
Proposition\,\ref{invconti} shows that 
the right hand side of \eqref{eq:form01} is  
changed on $\mathbb C{\setminus}\{-({\mathbb N}{+}\frac{1}{2})\}$ into  
$$ 
e_*^{(\log n{-}(1{+}\frac{1}{2}{+}
\cdots{+}\frac{1}{n}))(z\pm\frac{1}{i\h}u{\ctt}v)} 
{*}(z\pm\frac{1}{i\h}u{\ctt}v)_{*\pm}^{-1}{*}
\prod_{k=1}^n{*}
\Big(
\big(1{+}\frac{z\pm\frac{1}{i\h}u{\ctt}v}{k}\big)_{*\pm}^{-1}
{*}e_*^{\frac{1}{k}(z\pm\frac{1}{i\h}u{\ctt}v)}\Big). 
$$
Note that 
$\lim_{n{\to}\infty}
e_*^{(\log n{-}(1{+}\frac{1}{2}{+}\cdots{+}\frac{1}{n}))
(z\pm\frac{1}{i\h}u{\ctt}v)}
{=}e_*^{-\gamma(z\pm\frac{1}{\h i}u{\ctt}v)}$ is obvious, where  
$\gamma$ is Euler's constant.  
Thus, Proposition\,\ref{loose} gives the uniform convergence of 

\begin{equation}
  \label{eq:prodinf}
\begin{aligned}
&{\varGamma}_*(z{+}\frac{1}{i\h}u{\ctt}v)=
e_*^{-\gamma(z{+}\frac{1}{i\h}u{\ctt}v)}
{*}(z{+}\frac{1}{i\h}u{\ctt}v)_{*+}^{-1}{*}
\prod_{k=1}^{\infty}{*}
\Big(\big(1{+}\frac{1}{k}
(z{+}\frac{1}{i\h}u{\ctt}v)\big)_{*+}^{-1}
  {*}e_*^{\frac{1}{k}(z{+}\frac{1}{i\h}u{\ctt}v)}\Big)\\
&{\varGamma}_*(z{-}\frac{1}{i\h}u{\ctt}v)=
e_*^{-\gamma(z{-}\frac{1}{i\h}u{\ctt}v)}
{*}(z{-}\frac{1}{i\h}u{\ctt}v)_{*-}^{-1}{*}
\prod_{k=1}^{\infty}{*}
\Big(\big(1{+}\frac{1}{k}
(z{-}\frac{1}{i\h}u{\ctt}v)\big)_{*-}^{-1}
{*}e_*^{\frac{1}{k}(z{-}\frac{1}{i\h}u{\ctt}v)}\Big)
\end{aligned}
\end{equation} 
on each compact subset in the domain $\{z;{\rm{Re}}\,z>{-}1/2\}$. This
is the infinite product formula on the domain
$\{z;{\rm{Re}}\,z>{-}1/2\}$ in generic ordered expressions. 

\subsection{Analytic continuation of the infinite product formula} 

Here we show that the infinite product formula \eqref{eq:prodinf}
holds on $\mathbb C{\setminus}\{-({\mathbb N}{+}\frac{1}{2})\}$.
In generic ordered expression, the double integral 
$$
\int_{-\infty}^{\infty}\int_{-\infty}^0 
e^{-e^s}e_*^{(s{+}s')(z{\pm}\frac{1}{i\h}u{\ctt}v)}dsds'
$$
converges to give an element of $H\!ol({\mathbb C}^2)$. As
$z{\pm}\frac{1}{i\h}u{\ctt}v$ is a $*$-polynomial 
$$
(z{\pm}\frac{1}{i\h}u{\ctt}v){*}\int_{-\infty}^{\infty}\!\int_{-\infty}^0 
e^{-e^s}e_*^{(s{+}s')(z{\pm}\frac{1}{i\h}u{\ctt}v)}dsds'{=}
\int_{-\infty}^{\infty}\int_{-\infty}^0 
e_*^{s'(z{\pm}\frac{1}{i\h}u{\ctt}v)}{*}(z{\pm}\frac{1}{i\h}u{\ctt}v){*}
e^{-e^s}e_*^{s(z{\pm}\frac{1}{i\h}u{\ctt}v)}
dsds'
$$
makes sense. As 
$(z{\pm}\frac{1}{i\h}u{\ctt}v){*}
e^{-e^s}e_*^{s(z{\pm}\frac{1}{i\h}u{\ctt}v)}
{=}e^{-e^s}\frac{d}{ds}e_*^{s(z{\pm}\frac{1}{i\h}u{\ctt}v)}$,
integrating by $ds$ first gives 
$$
\int_{-\infty}^0 
e_*^{s'(z{\pm}\frac{1}{i\h}u{\ctt}v)}ds'{*}
\int_{-\infty}^{\infty}e^{-e^s}e_*^{s(z{+}1{\pm}\frac{1}{i\h}u{\ctt}v)}ds=
(z{\pm}\frac{1}{i\h}u{\ctt}v)_{*+}^{-1}{*}{\varGamma}_*(z{+}1{\pm}\frac{1}{i\h}u{\ctt}v)
$$
by the integration by parts. As 
$e_*^{s'(z{\pm}\frac{1}{i\h}u{\ctt}v)}{*}(z{\pm}\frac{1}{i\h}u{\ctt}v)
{=}\frac{d}{ds'}e_*^{s'(z{\pm}\frac{1}{i\h}u{\ctt}v)}$,  
 integrating by $ds'$ first gives 
$$
{\varGamma}_*(z{\pm}\frac{1}{i\h}u{\ctt}v)=
(z{\pm}\frac{1}{i\h}u{\ctt}v)_{*+}^{-1}{*}{\varGamma}_*(z{+}1{\pm}\frac{1}{i\h}u{\ctt}v),
\quad {\rm{Re}}\,z>{-}1/2.
$$
Recall that ${\varGamma}_*(z{\pm}\frac{1}{i\h}u{\ctt}v)$ and  
$(z{\pm}\frac{1}{i\h}u{\ctt}v)_{*+}^{-1}$ are analytically continued 
on ${\rm{Re}}\,z>-\frac{3}{2}$, $z\not=-\frac{1}{2}$ and 
${\varGamma}_*(z{+}1{\pm}\frac{1}{i\h}u{\ctt}v)$ is analytic on 
${\rm{Re}}\,z>-\frac{3}{2}$. 
By this we define 
$$
{\varGamma}_*(z{\pm}\frac{1}{i\h}u{\ctt}v)=
(z{\pm}\frac{1}{i\h}u{\ctt}v)_{*+}^{-1}{*}{\varGamma}_*(z{+}1{\pm}\frac{1}{i\h}u{\ctt}v),
\quad {\rm{Re}}\,z>{-}3/2, z\not=-\frac{1}{2}.
$$
Applying the infinite product formula \eqref{eq:prodinf} 
to the term ${\varGamma}_*(z{+}1{\pm}\frac{1}{i\h}u{\ctt}v)$
by replacing $z$ by $z{+}1$, and noting that 
$$
e^{-\gamma}\prod_{k=1}^{\infty}(1{-}\frac{1}{k})e^{\frac{1}{k}}=1,
$$ 
we see that the formula \eqref{eq:prodinf} holds for ${\rm{Re}}\,z>{-}3/2,
z\not=-\frac{1}{2}$. Repeating this procedure we see 
that the formula \eqref{eq:prodinf} holds on
$\mathbb C{\setminus}\{-({\mathbb N}{+}\frac{1}{2})\}$.

\begin{prop}\label{111}
In generic ordered expression,  
$$
\lim_{N\to\infty}
e_*^{\gamma(z{\pm}\frac{1}{i\h}u{\ctt}v)}
{*}(z{\pm}\frac{1}{i\h}u{\ctt}v){*}
\prod_{k=1}^{N}{*}
\Big(\big(1{+}\frac{1}{k}
(z{\pm}\frac{1}{i\h}u{\ctt}v)\big)
  {*}e_*^{-\frac{1}{k}(z{\pm}\frac{1}{i\h}u{\ctt}v)}\Big)
{*}{\varGamma}_*(z{\pm}\frac{1}{i\h}u{\ctt}v)=1.
$$
Namely, ${\varGamma}_*(z{\pm}\frac{1}{i\h}u{\ctt}v)$ is invertible 
on ${\mathbb C}{\setminus}\{-(\mathbb N{+}\frac{1}{2})\}$ in a certain 
weak sense. 
\end{prop}
Note that the proposition above does not necessarily imply 
the convergence of 
$$
\lim_{N\to\infty}
e_*^{\gamma(z{\pm}\frac{1}{i\h}u{\ctt}v)}
{*}(z{\pm}\frac{1}{i\h}u{\ctt}v){*}
\prod_{k=1}^{N}{*}
\Big(\big(1{+}\frac{1}{k}
(z{\pm}\frac{1}{i\h}u{\ctt}v)\big)
  {*}e_*^{-\frac{1}{k}(z{\pm}\frac{1}{i\h}u{\ctt}v)}\Big)
$$
in $H\!ol({\mathbb C}^2)$.  
In spite of this, it is still useful to
generalize as follows:

\begin{prop}\label{generalprop}
If $\varGamma_*(z{+}\frac{1}{i\h}u{\ctt}v){*}H_*$ is defined by
$$
\lim_{N\to\infty}
e_*^{{-}\gamma(z{\pm}\frac{1}{i\h}u{\ctt}v)}
{*}(z{\pm}\frac{1}{i\h}u{\ctt}v)_{*\pm}^{-1}{*}
\prod_{k=1}^{N}{*}
\Big(\big(1{+}\frac{1}{k}
(z{\pm}\frac{1}{i\h}u{\ctt}v)\big)_{*\pm}^{-1}
  {*}e_*^{\frac{1}{k}(z{\pm}\frac{1}{i\h}u{\ctt}v)}\Big){*}H_*,
$$
then 
$$
H_*{=}\lim_{N\to\infty}
e_*^{\gamma(z{\pm}\frac{1}{i\h}u{\ctt}v)}
{*}(z{\pm}\frac{1}{i\h}u{\ctt}v){*}
\prod_{k=1}^{N}{*}
\Big(\big(1{+}\frac{1}{k}
(z{\pm}\frac{1}{i\h}u{\ctt}v)\big)
  {*}e_*^{-\frac{1}{k}(z{\pm}\frac{1}{i\h}u{\ctt}v)}\Big)
{*}({\varGamma}_*(z{\pm}\frac{1}{i\h}u{\ctt}v){*}H_*)
$$
\end{prop}

\subsection{About Euler's reflection formula}
It is wellknown 
$$
\frac{\pi}{\sin\pi z}=
z^{-1}\prod_{n=1}^{\infty}(1{-}\frac{z^2}{n^2})^{-1}
=\varGamma(z)\varGamma(1{-}z).
$$
By the diagonal matrix expression, we have 
$$
{:}\frac{1}{\pi}\sin_*\pi(z{+}\frac{1}{i\h}u{\ctt}v){:}_{_{E(K)mat}}
{=}\sum_{k=0}^{\infty}\frac{1}{\pi}sin(\pi(z{+}k{+}\frac{1}{2}))E_{k,k}(K)
{=}\sum_{k=0}^{\infty}\frac{1}{\varGamma(z{+}k{+}\frac{1}{2})\varGamma(1{-}z{-}k{-}\frac{1}{2})}E_{k,k}(K).
$$
By the calculation as diagonal matrices, 
the r.h.s. may be replaced by 
$$
\Big(\sum_{k=0}^{\infty}\frac{1}{\varGamma(z{+}k{+}\frac{1}{2})}E_{k,k}(K)\Big){*}
\Big(\sum_{k=0}^{\infty}\frac{1}{\varGamma(1{-}z{-}k{-}\frac{1}{2})}E_{k,k}(K)\Big).
$$
By \eqref{totresult}, we see the first component converges in
$H\!ol({\mathbb C}^2)$, but the second component diverges in
$H\!ol({\mathbb C}^2)$. Note also that 
$\sum_{k=0}^{\infty}
\frac{1}{\varGamma(1{-}z{-}k{-}\frac{1}{2})}{\overline E}_{k,k}(K)$
converges.

Suggested by Theorem\,\ref{invGamma} and
$E_{k,k}(K){*}\overline{E}_{l.l}(K){=}0$, 
we have the following  

\bigskip
\noindent
{\bf Conjecture}\,\,There is no expression parameter such that
$$
\varGamma_*^{-1}(z{+}\frac{1}{i\h}u{\ctt}v){*}\varGamma_*^{-1}(1{-}z{-}\frac{1}{i\h}u{\ctt}v)
$$
is defined in $H\!ol({\mathbb C}^2)$. This implies that Euler's
reflection formula fails for $\sin_*\pi(z{+}\frac{1}{i\h}u{\ctt}v)$.

\section{Star-zeta function} 
As it is wellknown, the gamma function and the zeta function are
deeply related. 
Ordinary zeta function $\zeta(s)$ is defined by 
$$
\zeta(s)=\sum_{n=1}^{\infty}(\frac{1}{n})^s, \quad {\rm{Re}}\,s>1.
$$
Replacing $1/n$ by $e_*^{-(\log n)(z{\pm}\frac{1}{i\h}u{\ctt}v)}$, 
we define the $*$-zeta function by 
\begin{equation}\label{star-zeta}
\zeta_*(z{\pm}\frac{1}{i\h}u{\ctt}v)=
\sum_{n=1}^{\infty}e_*^{-(\log n)(z{\pm}\frac{1}{i\h}u{\ctt}v)}.
\end{equation}
Comparing this with the integral  
$\int_1^{\infty}e_*^{-(\log t)(z{\pm}\frac{1}{i\h}u{\ctt}v)}dt$  
in generic ordered expression and replacing $t=e^s$, we have a remarkable 
\begin{prop}\label{remarkable}
In generic ordered expression, 
$\zeta_*(z{\pm}\frac{1}{i\h}u{\ctt}v)$ absolutely converges uniformly on every
compact domain in $\{z; {\rm{Re}}\,z{>}1/2\}\times{\mathbb C}^2$.
\end{prop}

\bigskip
Let $p_1<p_2< p_3<\cdots$ be 
the series of all prime numbers; 
i.e. $p_1{=}2$, $p_2{=}3$,  $p_3{=}5$, $\cdots$. 
Set 
$$
1-p_n^{-(z{\pm}\frac{1}{i\h}u{\ctt}v)_*}=1-e_*^{-(z{\pm}\frac{1}{i\h}u{\ctt}v)\log p_n}, \quad 
(1-e_*^{-(z{\pm}\frac{1}{i\h}u{\ctt}v)\log p_n})_{*+}^{-1}=
\sum_{k=0}^{\infty}e_*^{-k(z{\pm}\frac{1}{i\h}u{\ctt}v)\log p_n}.
$$
The latter converges in generic ordered expression. 
Thus, in generic ordered expression  
$$
\prod_{n=1}^{\ell}(1-e_*^{-(z{\pm}\frac{1}{i\h}u{\ctt}v)\log p_n})_*^{-1}
$$
is welldefined to obtain  
$$
\begin{aligned}
\sum e_*^{-k_1(z{\pm}\frac{1}{i\h}u{\ctt}v)\log p_1}
{*}e_*^{-k_2(z{\pm}\frac{1}{i\h}u{\ctt}v)\log p_2}
&{*}\cdots
{*}e_*^{-k_{\ell}(z{\pm}\frac{1}{i\h}u{\ctt}v)\log p_\ell}\\
&=
\sum_{k_1,k_2,\dots,k_{\ell}\geq 0}
e_*^{-(z{\pm}\frac{1}{i\h}u{\ctt}v)
(\log p_1^{k_1}p_2^{k_2}\cdots p_\ell^{k_{\ell}})}.
\end{aligned}
$$
Thus, we have 
\begin{equation}\label{Eulerprod00}
\lim_{\ell\to\infty}
\prod_{n=1}^{\ell}(1-e_*^{-(z{\pm}\frac{1}{i\h}u{\ctt}v)\log p_\ell})_*^{-1}
=
\sum_{k_1,k_2,\dots,k_{\ell}\geq 0}
e_*^{-(z{\pm}\frac{1}{i\h}u{\ctt}v)
(\log p_1^{k_1}p_2^{k_2}\cdots p_\ell^{k_{\ell}})}.
\end{equation}
Hence taking $\ell\to\infty$, 
the summation runs through all positive integers once for all 
by the unique factorization theorem for natural numbers. 
Thus, we have the  Euler's infinite product formula 
\begin{equation}\label{Eulerprod00}
\lim_{\ell\to\infty}
\prod_{n=1}^{\ell}
(1-e_*^{-(z{\pm}\frac{1}{i\h}u{\ctt}v)\log p_n})_{*}^{-1}
=\sum_{n=1}^{\infty}e_*^{-(z{\pm}\frac{1}{i\h}u{\ctt}v)\log n}=
\zeta_*(z{\pm}\frac{1}{i\h}u{\ctt}v).
\end{equation}

\bigskip
On the other hand, 
$$
\begin{aligned}
\prod_{n=1}^{\ell}&(1-e_*^{-(z{+\pm}\frac{1}{i\h}u{\ctt}v)\log p_n})=\\
&
1{-}\sum_{n=1}^{\infty}e_*^{-(z{\pm}\frac{1}{i\h}u{\ctt}v)\log p_n}{+}\!\!
\sum_{1{\leq}i{<}j\leq \ell}e_*^{-(z{\pm}\frac{1}{i\h}u{\ctt}v)\log p_ip_j}
{-}\cdots{+}
(-1)^{\ell}e_*^{-(z{\pm}\frac{1}{i\h}u{\ctt}v)\log p_1p_2\cdots p_\ell}.
\end{aligned}
$$
Note that $p_{i_1}p_{i_2}\cdots p_{i_k}$ which appear 
in the summation are much less than the positive integers.
Hence, the r.h.s. absolutely converges uniformly on every
 compact region of $z{\pm}\frac{1}{i\h}u{\ctt}v$ such that ${\rm{Re}}\,z>\frac{1}{2}$. 
Hence, we have 
\begin{prop}\label{invzeta}
In generic $K$-expression, both
${:}\zeta_*(z{\pm}\frac{1}{i\h}u{\ctt}v){:}_{_K}$ and 
${:}\zeta_*(z{\pm}\frac{1}{i\h}u{\ctt}v)_{*}^{-1}{:}_{_K}$ exist on 
the domain ${\rm{Re}}\,z>\frac{1}{2}$. 
\end{prop}

\subsection{Generating function of numbers of partitions}

Note that $*$-zeta function is still useful to treat prime numbers. 
Note that the mathematics is not based on the notion of cardinal numbers but 
on the notion of natural numbers axiomatized by the notion of 
ordinal numbers. 
Our ultimate theme is to know how mathematics recognize
cardinal numbers and the ``time''. 

Although this is not directly relevant to our purpose, a similar 
formula is proposed relating to the cardinal numbers. 

Consider the function 
$$
f(x)=
1{+}\sum_{n=1}^{\infty}p(n)x^n{=}\prod_{n=1}^{\infty}(1{-}x^n)^{-1},
\quad |x|<1. 
$$
Replacing $x$ by $e_*^{{-}(z{\pm}\frac{1}{i\h}u{\ctt}v)}$, we define 
$$
f_*(e_*^{z{\pm}\frac{1}{i\h}u{\ctt}v})=
\prod_{n=1}^{\infty}{*}(1{-}e_*^{-n(z{\pm}\frac{1}{i\h}u{\ctt}v)})^{-1}
$$
Note that if ${\rm{Re}}\,z\geq 0$, then in generic ordered expression 
$$
(1{-}e_*^{-n(z{\pm}\frac{1}{i\h}u{\ctt}v)})^{-1}=
\sum_{k=0}^{\infty}e_*^{-k\,n(z{\pm}\frac{1}{i\h}u{\ctt}v)}
$$
converges absolutely and uniformly on every
 compact region of $z{\pm}\frac{1}{i\h}u{\ctt}v$ such that 
 $\rm{Re}\,z>0$, 
 and $f_*(e_*^{z{\pm}\frac{1}{i\h}u{\ctt}v})$ plays the same
role as a generating function of the number of partitions $p(n)$.
$$
1{+}\sum_{n=1}^{\infty}p(n)e_*^{-n(z{\pm}\frac{1}{i\h}u{\ctt}v)}{=}
\prod_{\ell=1}^{\infty}{*}\sum_{k=0}^{\infty}e_*^{-k\ell(z{\pm}\frac{1}{i\h}u{\ctt}v)}.
$$
This means that the notion of cardinal number is independent of
algebraic structure.

\bigskip
It is known by Postnikov that 
$$
p(n)=\frac{\exp(\pi\sqrt{2n/3})}{4\sqrt{3}n}\Big(1{+}O(\frac{\log n}{n^{1/4}})\Big).
$$
Hence  $f_*(e_*^{z{\pm}\frac{1}{i\h}u{\ctt}v})$ converges for ${\rm{Re}}\,z>-\frac{1}{2}$.
By the same reason as in $\zeta_*(z{\pm}\frac{1}{i\h}u{\ctt}v)$ we have 

\begin{prop}\label{invpart00}
In generic ordered expression $K$, both  
${:}f_*(z{\pm}\frac{1}{i\h}u{\ctt}v){:}_{_K}$ and
${:}f_*(e_*^{z{\pm}\frac{1}{i\h}u{\ctt}v})_*^{-1}{:}_{_K}$ exist on 
the domain $\rm{Re}\,z>0$. 
\end{prop}

\bigskip 
\noindent
{\bf Note}\,\,\,There is no analytic continuation of the generating
function $f_*(z{+}\frac{1}{i\h}u{\ctt}v)$ of number of partitions. It
is very likely that ${\rm{Re}}\,z=0$ is the natural boundary.

\subsection{Analytic continuation of
  $\zeta_*(z{\pm}\frac{1}{i\h}u{\ctt}v)$}

First of all, note that the next equality holds for every $a\in \mathbb R$ and for every
$*$-polynomial $p_*(u,v)$ 
$$
p_*(u,v){*}e_*^{a(z{\pm}\frac{1}{i\h}u{\ctt}v)}
{*}\varGamma_*(z\pm\frac{1}{i\h}u{\ctt}v){=}
\int_{-\infty}^{\infty}e^{-e^s}
p_*(u,v){*}e_*^{(s{+}a)(z\pm\frac{1}{i\h}u{\ctt}v)}ds.
$$
Now, set $e^s=ne^{\sigma}$ in the definition of $*$-gamma function, we have 
$$
\varGamma_*(z\pm\frac{1}{i\h}u{\ctt}v)=
\int_{-\infty}^{\infty}e^{-e^s}
e_*^{s(z\pm\frac{1}{i\h}u{\ctt}v)}ds
=
\int_{-\infty}^{\infty}e^{-ne^{\sigma}}
e_*^{(\sigma+\log n)(z{\pm}\frac{1}{i\h}u{\ctt}v)}d\sigma.
$$
The exponential law 
$e_*^{(\sigma+\log n)(z{\pm}\frac{1}{i\h}u{\ctt}v)}
{=}e_*^{\sigma(z{\pm}\frac{1}{i\h}u{\ctt}v)}{*}
e_*^{(\log n)(z{\pm}\frac{1}{i\h}u{\ctt}v)}$  
and the continuity of  
$e_*^{-(\log n)(z{\pm}\frac{1}{i\h}u{\ctt}v)}{*}$ 
give 
$$
\varGamma_*(z{\pm}\frac{1}{i\h}u{\ctt}v){*}
e_*^{-(z{\pm}\frac{1}{i\h}u{\ctt}v)\log n}=
\int_{-\infty}^{\infty}e^{-ne^{\sigma}}
e_*^{\sigma(z{\pm}\frac{1}{i\h}u{\ctt}v)}d\sigma.
$$
It follows
\begin{equation}\label{beforlim}
\sum_{n=1}^{N}e_*^{-(z{\pm}\frac{1}{i\h}u{\ctt}v)\log n}=
\varGamma_*(z{\pm}\frac{1}{i\h}u{\ctt}v)^{-1}
{*}\int_{-\infty}^{\infty}\sum_{n=1}^Ne^{-ne^{\sigma}}
e_*^{\sigma(z{\pm}\frac{1}{i\h}u{\ctt}v)}d\sigma.
\end{equation}
where $\varGamma_*(z{\pm}\frac{1}{i\h}u{\ctt}v)^{-1}$ is 
defined by Theorem\,\,\ref{invGamma}.

\medskip
By taking the limit $N\to\infty$, 
the l.h.s. converges easily in $H{\!o}l({\mathbb C}^2)$ 
to obtain for every $a\in \mathbb C$ and for every
$*$-polynomial $p_*(u,v)$ 
\begin{equation}\label{zetainteg}
\begin{array}{l}
\qquad p_*(u,v){*}e_*^{a(z{\pm}\frac{1}{i\h}u{\ctt}v)}
{*}\zeta_*(z{\pm}\frac{1}{i\h}u{\ctt}v)\\
=
 p_*(u,v){*}e_*^{a(z{\pm}\frac{1}{i\h}u{\ctt}v)}
{*}\varGamma_*(z{\pm}\frac{1}{i\h}u{\ctt}v)^{-1}
{*}\int_{-\infty}^{\infty}\frac{1}{e^{e^{\sigma}}{-}1}
e_*^{\sigma(z{\pm}\frac{1}{i\h}u{\ctt}v)}d\sigma.
\end{array}
\end{equation}
 
Consider the main part
$\int_{-\infty}^{\infty}\frac{1}{e^{e^{\sigma}}{-}1}
e_*^{\sigma(z{\pm}\frac{1}{i\h}u{\ctt}v)}d\sigma$ 
of the r.h.s. of \eqref{zetainteg}. 
By the definition of the Bernoulli numbers, we have 
$$
\frac{1}{e^{e^t}{-}1}=
\frac{1}{e^{t}}-\frac{1}{2}+
\sum_{n\geq 1}B_{2n}\frac{(-1)^{n-1}}{(2n)!}e^{(2n-1)t},
\quad B_{2n+1}=0, \,\, B_{2n}>0. 
$$
Considering domains $t>0$ and $t<0$ separately, we see 
$B(t)=\frac{1}{e^{e^t}{-}1}{-}e^{-t}$
is rapidly decreasing function.

\bigskip

By Proposition\,\,\ref{generalprop} and \eqref{zetainteg}, we get  
\begin{prop}\label{elimfactor}
$(z{+}\frac{1}{i\h}u{\ctt}v){*}\prod_{\ell=1}^n{*}
(1{+}\frac{1}{\ell}(z{+}\frac{1}{i\h}u{\ctt}v)){*}
e_*^{-\frac{1}{\ell}(z{+}\frac{1}{i\h}u{\ctt}v)}
{*}\int_{-\infty}^{\infty}
\frac{1}{e^{e^{\sigma}}{-}1}
e_*^{\sigma(z{+}\frac{1}{i\h}u{\ctt}v)}d\sigma$ is holomorphic on the
domain 
${\mathbb C}{\setminus}
\{-({\mathbb N}{+}n{+}\frac{1}{2})\cup\{\frac{1}{2}\}$. Taking
$n\to\infty$ we have  
$$
\begin{aligned}
&\lim_{n\to\infty}e_*^{\gamma(z{+}\frac{1}{i\h}u{\ctt}v))}
{*}(z{+}\frac{1}{i\h}u{\ctt}v){*}\prod_{\ell=1}^n{*}
(1{+}\frac{1}{\ell}(z{+}\frac{1}{i\h}u{\ctt}v)){*}
e_*^{-\frac{1}{\ell}(z{+}\frac{1}{i\h}u{\ctt}v)}
{*}\varGamma_*(z{+}\frac{1}{i\h}u{\ctt}v){*}
\zeta_*(z{+}\frac{1}{i\h}u{\ctt}v)\\
&=
\lim_{n\to\infty}e_*^{\gamma(z{+}\frac{1}{i\h}u{\ctt}v))}{*}
(z{+}\frac{1}{i\h}u{\ctt}v){*}\prod_{\ell=1}^n{*}
(1{+}\frac{1}{\ell}(z{+}\frac{1}{i\h}u{\ctt}v)){*}
e_*^{-\frac{1}{\ell}(z{+}\frac{1}{i\h}u{\ctt}v)}
{*}\int_{-\infty}^{\infty}
\frac{1}{e^{e^{\sigma}}{-}1}
e_*^{\sigma(z{+}\frac{1}{i\h}u{\ctt}v)}d\sigma.
\end{aligned}
$$ 
Since the first line converges to
$\zeta_*(z{+}\frac{1}{i\h}u{\ctt}v)$, we see
$\zeta_*(z{+}\frac{1}{i\h}u{\ctt}v)$ 
is holomorphic on ${\mathbb C}{\setminus}\{\frac{1}{2}\}$ in generic
ordered expression.
Similarly, $\zeta_*(z{-}\frac{1}{i\h}u{\ctt}v)$ 
is holomorphic on ${\mathbb C}{\setminus}\{\frac{1}{2}\}$ in generic
ordered expression.
\end{prop}

By the proof of Lemma\,\ref{analcont}, we see that the residue of 
$\zeta_*(z{+}\frac{1}{i\h}u{\ctt}v)$ is given by 
$$
{\rm{Res}}(\zeta_*(z{+}\frac{1}{i\h}u{\ctt}v), z{=}\frac{1}{2})=
{\rm{Res}}((z{+}\frac{1}{i\h}u{\ctt}v)_{*-}^{-1}, 
z{=}\frac{1}{2}) ={\overline{\varpi}}_{00}.
$$
where ${\overline{\varpi}}_{00}$ in the r.h.s. is called 
the bar-vacuum.

However, in the proof of Proposition\,\ref{elimfactor}, we have only
to eliminate the singularities of \\
$\int_{-\infty}^{\infty}
\frac{1}{e^{e^{\sigma}}{-}1}
e_*^{\sigma(z{+}\frac{1}{i\h}u{\ctt}v)}d\sigma$ 
by multiplying odd factors 
$$
(z{+}\frac{1}{i\h}u{\ctt}v){*}\prod_{\ell=1}^{n}{*}
(1{+}\frac{1}{2\ell{+}1}(z{+}\frac{1}{i\h}u{\ctt}v)){*}
e_*^{-\frac{1}{2\ell{+}1}(z{+}\frac{1}{i\h}u{\ctt}v)}{*}.
$$ 
Even factors $\prod_{\ell=1}^{n}
(1{+}\frac{1}{2\ell}(z{+}\frac{1}{i\h}u{\ctt}v)){*}
e_*^{-\frac{1}{\ell}(z{+}\frac{1}{i\h}u{\ctt}v)}{*}$
are used only to get gamma function. Thus,
corresponding to the trivial zero's of ordinary Riemann zeta function,
we see that $\zeta_*(z{+}\frac{1}{i\h}u{\ctt}v)$ has many factors: For
every positive integer $n$, 
$$
\zeta_*(z{+}\frac{1}{i\h}u{\ctt}v)=(z{+}2n{+}\frac{1}{i\h}u{\ctt}v){*}g_n(z{+}\frac{1}{i\h}u{\ctt}v)
$$
where $g_n(z{+}\frac{1}{i\h}u{\ctt}v)$ is holomorphic on 
${\mathbb C}{\setminus}\{\frac{1}{2}\}$.
Thus, Proposition\,\ref{elimfactor} does not imply that 
$\zeta_*(z{\pm}\frac{1}{i\h}u{\ctt}v)_*^{-1}$ is holomorphic on 
${\mathbb C}{\setminus}\{\frac{1}{2}\}$, but many singular points
on negative real axis.

\subsubsection{Diagonal matrix expression of $\zeta_*(z{\pm}\frac{1}{i\h}u{\ctt}v)$}
First, we show 
\begin{lem}\label{Lemmahol}
In a generic ordered expression
$$
p_*(u,v){*}e_*^{a(z{\pm}\frac{1}{i\h}u{\ctt}v)}
{*}\int_{-\infty}^{\infty}
\frac{1}{e^{e^{\sigma}}{-}1}
e_*^{\sigma(z{\pm}\frac{1}{i\h}u{\ctt}v)}d\sigma 
$$ 
is a $H\!ol({\mathbb C}^2)$-valued 
holomorphic function of $z$ on the domain ${\rm{Re}}\,z{>}\frac{1}{2}$
for every $a\in \mathbb C$ and for every $*$-polynomial $p_*(u,v)$.  
\end{lem} 

\noindent
{\bf Proof}\,\,It is clear that  
$\int_{0}^{\infty}
\frac{1}{e^{e^{\sigma}}{-}1}
e_*^{\sigma(z{\pm}\frac{1}{i\h}u{\ctt}v)}d\sigma$ 
is a $H\!ol({\mathbb C}^2)$-valued entire function of $z$. 

For 
$\int_{-\infty}^{0}
\frac{1}{e^{e^{\sigma}}{-}1}
e_*^{\sigma(z{\pm}\frac{1}{i\h}u{\ctt}v)}d\sigma$, note  
that $e^{e^{\sigma}}-1>e^{\sigma}$ and 
$e_*^{\sigma \frac{1}{i\h}u{\ctt}v}$ is of order 
$e^{-|\sigma|^{\frac{1}{2}}}$ in generic ordered expressions. \hfill$\Box$

The diagonal matrix expressions of 
$\int_{0}^{\infty}
\frac{1}{e^{e^{\sigma}}{-}1}
e_*^{\sigma(z{\pm}\frac{1}{i\h}u{\ctt}v)}d\sigma$ are 
\begin{equation}\label{diagexp}
\begin{aligned}
{:}\int_{0}^{\infty}
\frac{1}{e^{e^{s}}{-}1}
e_*^{s(z{+}\frac{1}{i\h}u{\ctt}v)}ds{:}_{_{E(K)mat}}
=&
\sum_{k=0}^{\infty}\int_{0}^{\infty}
\frac{1}{e^{e^{s}}{-}1}e^{s(z{+}n{+}\frac{1}{2})}ds E_{n,n}(K)\\
{:}\int_{0}^{\infty}
\frac{1}{e^{e^{s}}{-}1}
e_*^{s(z{-}\frac{1}{i\h}u{\ctt}v)}ds{:}_{_{\overline{E}(K)mat}}
=&
\sum_{k=0}^{\infty}\int_{0}^{\infty}
\frac{1}{e^{e^{\sigma}}{-}1}e^{s(z{-}n{-}\frac{1}{2})}ds \overline{E}_{n,n}(K)
\end{aligned}
\end{equation}
and every component is an entire function of $z$.

\begin{lem}\label{analcont}
$\int_{-\infty}^{\infty}
\frac{1}{e^{e^{\sigma}}{-}1}
e_*^{\sigma(z{+}\frac{1}{i\h}u{\ctt}v)}d\sigma$  
is analytically continued to the domain 
${\mathbb C}{\setminus}
\{-({\mathbb N}{+}\frac{1}{2})\cup\{\frac{1}{2}\}\}$. Similarly, 
$\int_{-\infty}^{\infty}
\frac{1}{e^{e^{\sigma}}{-}1}
e_*^{\sigma(z{-}\frac{1}{i\h}u{\ctt}v)}d\sigma$  
is analytically continued to the domain 
${\mathbb C}{\setminus}
\{-({\mathbb N}{+}\frac{1}{2})\cup\{\frac{1}{2}\}\}$.

We denote these by 
$L_*(z{+}\frac{1}{i\h}u{\ctt}v)$, $L_*(z{-}\frac{1}{i\h}u{\ctt}v)$ respectively. 
\end{lem}

\noindent
{\bf Proof}\,\,It is enough to show for the integral 
$\int_{-\infty}^{0}
\frac{1}{e^{e^{\sigma}}{-}1}
e_*^{\sigma(z{\pm}\frac{1}{i\h}u{\ctt}v)}d\sigma$.  

For every fixed $z$ find an positive integer $N$ such 
that ${\rm{Re}}z{+}N>\frac{1}{2}$, and set 
$$ 
\frac{1}{e^{e^t}{-}1}=
\frac{1}{e^{t}}-\frac{1}{2}+\cdots+
B_{2N}\frac{(-1)^{N-1}}{(2N)!}e^{(2N-1)t}+
e^{(2N-1)t}R_n(t).
$$
It is easy to see that 
$$
\int_{-\infty}^{0}e^{(2N-1)t}R_n(t){*}e_*^{\sigma(z{+}\frac{1}{i\h}u{\ctt}v)}d\sigma
$$
converges to give a holomorphic function defined on the domain ${\rm{Re}}(z{+}N)>\frac{1}{2}$.  
It follows by termwise integration 
$$
\begin{aligned}
\int_{-\infty}^{0}&
\frac{1}{e^{e^{t}}{-}1}
e_*^{t(z{+}\frac{1}{i\h}u{\ctt}v)}dt
=
\int_{-\infty}^{0}
\big(e^{-t}{-}\frac{1}{2}
{+}\sum_{n\geq 0}(-1)^{n-1}B_{2n}\frac{e^{(2n-1)t}}{(2n)!}\big)
e_*^{t(z{+}\frac{1}{i\h}u{\ctt}v)}dt\\
&=
((z{-}1){+}\frac{1}{i\h}u{\ctt}v)^{-1}_{*+}
{-}\frac{1}{2}(z{+}\frac{1}{i\h}u{\ctt}v)^{-1}_{*+}
{+}\sum_{n\geq 0}(-1)^{n-1}\frac{B_{2n}}{(2n)!}
(z{+}2n{-}1{+}\frac{1}{i\h}u{\ctt}v)_{*+}^{-1}.
\end{aligned}
$$
Similarly, 
$$
\begin{aligned}
\int_{-\infty}^{0}&
\frac{1}{e^{e^{t}}{-}1}
e_*^{t(z{-}\frac{1}{i\h}u{\ctt}v)}ds
=
\int_{-\infty}^{0}
\big(e^{-t}{-}\frac{1}{2}
{+}\sum_{n\geq 0}(-1)^{n-1}B_{2n}\frac{e^{(2n-1)t}}{(2n)!}\big)
e_*^{t(z{-}\frac{1}{i\h}u{\ctt}v)}dt \\
&=
((z{-}1){-}\frac{1}{i\h}u{\ctt}v)^{-1}_{*-}
{-}\frac{1}{2}(z{-}\frac{1}{i\h}u{\ctt}v)^{-1}_{*-}
{+}\sum_{n\geq 0}(-1)^{n-1}\frac{B_{2n}}{(2n)!}
(z{+}2n{-}1{-}\frac{1}{i\h}u{\ctt}v))_{*-}^{-1}\\
&=
{-}(-(z{-}1){+}\frac{1}{i\h}u{\ctt}v)^{-1}_{*-}
{+}\frac{1}{2}(-z{+}\frac{1}{i\h}u{\ctt}v)^{-1}_{*-}
{-}\sum_{n\geq 0}(-1)^{n-1}\frac{B_{2n}}{(2n)!}
(-(z{+}2n{-}1){+}\frac{1}{i\h}u{\ctt}v))_{*-}^{-1}
\end{aligned}
$$
The result follows from the analytic continuations 
of $(z{+}2n{-}1{\pm}\frac{1}{i\h}u{\ctt}v))_{*\pm}^{-1}$ 
given in  \cite{ommy6}. （Cf. Theorem\,\ref{deltacont} also).
${}$\hfill$\Box$

\bigskip
By Lemmas\,\ref{Lemmahol} and \ref{analcont}, we see that  
in a generic ordered expression,
$$ 
\varGamma_*(z{\pm}\frac{1}{i\h}u{\ctt}v){*}
\zeta_*(z{\pm}\frac{1}{i\h}u{\ctt}v)=
\int_{-\infty}^{\infty}
\frac{1}{e^{e^{\sigma}}{-}1}
e_*^{\sigma(z{+}\frac{1}{i\h}u{\ctt}v)}d\sigma
$$ 
is a $H\!ol({\mathbb C}^2)$-valued holomorphic function of 
$z$ defined on ${\rm{Re}}\,z>\frac{1}{2}$, 
and this is continued to the domain 
${\mathbb C}{\setminus}
\{-({\mathbb N}{+}\frac{1}{2})\cup\{\frac{1}{2}\}\}$ 
analytically. 

The $E(K)mat$ and ${\overline{E}(K)mat}$-expressions of 
$\int_{-\infty}^{\infty}\frac{1}{e^{e^{\sigma}}{-}1}e_*^{\sigma(z{\pm}\frac{1}{i\h}u{\ctt}v)}d\sigma$
are obtained by these formulas by splitting the integral into 
$\int_{0}^{\infty}{+}\int_{-\infty}^{0}$.
We denote these as follows:
\begin{equation}\label{matLpm}
\begin{aligned}
&{:}L_*(z{+}\frac{1}{i\h}u{\ctt}v){:}_{_{E(K)mat}}=
\sum_{k=0}^{\infty}L(z{+}k{+}\frac{1}{2})E_{k,k}(K)\\
&{:}L_*(z{-}\frac{1}{i\h}u{\ctt}v){:}_{_{\overline{E}(K)mat}}= 
\sum_{k=0}^{\infty}L(z{+}k{+}\frac{1}{2})\overline{E}_{k,k}(K).
\end{aligned}
\end{equation}
$L(z{+}k{+}\frac{1}{2})$ are holomorphic on ${\mathbb C}{\setminus}
\{-({\mathbb N}{+}\frac{1}{2})\cup\{\frac{1}{2}\}\}$.

Combining this with \eqref{totresult} as diagonal matrix calculations, we have  
\begin{equation}\label{matzeta}
\begin{aligned}
&{:}\zeta_*(z{+}\frac{1}{i\h}u{\ctt}v){:}_{_{E(K)mat}}{=}
\sum_{k=0}^{\infty}\zeta(z{+}k{+}\frac{1}{2})E_{k,k}(K){=}
\sum_{k=0}^{\infty}\varGamma^{-1}(z{+}k{+}\frac{1}{2})L(z{+}k{+}\frac{1}{2})E_{k,k}(K)\\
&{:}\zeta_*(z{-}\frac{1}{i\h}u{\ctt}v){:}_{_{\overline{E}(K)mat}}{=}
\sum_{k=0}^{\infty}\zeta(z{+}k{+}\frac{1}{2})\overline{E}_{k,k}(K){=}
\sum_{k=0}^{\infty}\varGamma^{-1}(z{+}k{+}\frac{1}{2})L(z{+}k{+}\frac{1}{2})\overline{E}_{k,k}(K).
\end{aligned}
\end{equation}

Since all singular point of $L(z{+}k{+}\frac{1}{2})$ except $1/2$ are
eliminated by $\varGamma^{-1}(z{+}k{+}\frac{1}{2})$, we have 
\begin{prop}\label{singsingzeta}
Both ${:}\zeta_*(z{+}\frac{1}{i\h}u{\ctt}v){:}_{_{E(K)mat}}$ and 
${:}\zeta_*(z{-}\frac{1}{i\h}u{\ctt}v){:}_{_{\overline{E}(K)mat}}$ are
holomorphic on ${\mathbb C}{\setminus}\{1/2\}$.
\end{prop}
In the next section, we prove this more generally.

\bigskip
By these we make the diagonal matrix expressions of their inverses:
\begin{equation}\label{matzetainv}
\begin{aligned}
{:}\zeta_*(z{+}\frac{1}{i\h}u{\ctt}v)^{-1}{:}_{_{E(K)mat}}=
\sum_{k=0}^{\infty}\zeta(z{+}k{+}\frac{1}{2})^{-1}E_{k,k}(K)\\
{:}\zeta_*(z{-}\frac{1}{i\h}u{\ctt}v)^{-1}{:}_{_{\overline{E}(K)mat}}= 
\sum_{k=0}^{\infty}\zeta(z{+}k{+}\frac{1}{2})^{-1}\overline{E}_{k,k}(K).
\end{aligned}
\end{equation}

\begin{prop}\label{zeroandsing}
If $\zeta(z_0{+}n{+}\frac{1}{2})=0$ for some $z_0{+}n$, then both 
$$
{:}\zeta_*(z_0{+}\frac{1}{i\h}u{\ctt}v)^{-1}{:}_{_{E(K)mat}},\quad 
{:}\zeta_*(z_0{-}\frac{1}{i\h}u{\ctt}v)^{-1}{:}_{_{\overline{E}(K)mat}}
$$
 must be singular. 
\end{prop}

\section{Reflection property of  
$\zeta_*(s{\pm}\frac{1}{i\h}u{\ctt}v)$}\label{func-idntty22}

Now, setting $e^s=n^2\pi e^{x}$ in the definition of 
$*$-gamma function, the exponential law and the continuity of   
the ${*}$-multiplication 
$e_*^{(s{+}\frac{1}{i\h}u{\ctt}v)(\log n{+}\frac{1}{2}\log\pi)}{*}$ 
give 
$$
\begin{aligned}
\varGamma_*(\frac{1}{2}(s{\pm}\frac{1}{i\h}u{\ctt}v))=&
\int_{-\infty}^{\infty}e^{-n^2\pi e^x}
e_*^{(2\log n{+}\log\pi{+}x)\frac{1}{2}
(s{\pm}\frac{1}{i\h}u{\ctt}v)}dx\\
=& 
e_*^{(s{\pm}\frac{1}{i\h}u{\ctt}v)
(\log n{+}\frac{1}{2}\log\pi)}{*}
\int_{-\infty}^{\infty} 
e^{-n^2\pi e^x} 
e_*^{\frac{1}{2}x(s{\pm}\frac{1}{i\h}u{\ctt}v)}dx,\quad {\rm{Re}}\,s>{-}\frac{1}{2}.
\end{aligned} 
$$
Hence, we have for every positive integer 
$$
\varGamma_*(\frac{1}{2}(s{\pm}\frac{1}{i\h}u{\ctt}v)){*}
e_*^{-\frac{1}{2}\log\pi(s{\pm}\frac{1}{i\h}u{\ctt}v)}{*}
e_*^{-(\log n)(s{\pm}\frac{1}{i\h}u{\ctt}v)}
{=}
\int_{-\infty}^{\infty} 
e^{-n^2\pi e^x} 
e_*^{\frac{1}{2}x(s{\pm}\frac{1}{i\h}u{\ctt}v)}dx,\quad {\rm{Re}}\,s>{-}\frac{1}{2}.
$$
It follows by checking associativity that 
$$
\varGamma_*(\frac{1}{2}(s{\pm}\frac{1}{i\h}u{\ctt}v)){*}
e_*^{-\frac{1}{2}(s{\pm}\frac{1}{i\h}u{\ctt}v)\log\pi}{*}
\zeta_*(s{\pm}\frac{1}{i\h}u{\ctt}v)=
\lim_{N\to\infty}
\int_{-\infty}^{\infty} 
\sum_{n=1}^Ne^{-n^2\pi e^x} 
e_*^{\frac{1}{2}x(s{\pm}\frac{1}{i\h}u{\ctt}v)}dx.
$$
Note that the term 
$\sum_{n=1}^{\infty}e^{-n^2\pi e^x}$ is rapidly decreasing 
in the positive direction w.r.t. $x$ and bounded in the negative direction. 
${:}e_*^{\frac{1}{2}x(s{\pm}\frac{1}{i\h}u{\ctt}v)}{:}_{_K}$ is 
rapidly decreasing in the both directions in generic ordered expressions.
Hence we see that the r.h.s. can be written as  
\begin{equation}\label{phiphi}
\Phi_*(s{\pm}\frac{1}{i\h}u{\ctt}v)=\int_{-\infty}^{\infty} 
\sum_{n=1}^{\infty}e^{-n^2\pi e^x} 
e_*^{\frac{1}{2}x(s{\pm}\frac{1}{i\h}u{\ctt}v)}dx
\end{equation}
and this is holomorphic on the domain 
${\rm{Im}}\,s>{-}\frac{1}{2}$, where 
$\sum_{n=1}^{\infty}e^{-n^2\pi e^x}$ in the integrand 
is the value at $w=0$ of Jacobi's theta function  
$\theta_3(w,\tau)$ given in \cite{OMMY3}, that is 
$$
\sum_{n=1}^{\infty}e^{-n^2\pi e^x}= 
\frac{1}{2}\big(\theta_3(0, \pi e^x)-1\big)
$$
and recall here the famous theta relation 
$\theta_3(0,{\tau})=
\sqrt{\frac{\pi}{\tau}}\theta_3(0,\frac{\pi^2}{\tau})$ 
given in \cite{OMMY3}. This gives in particular 
\begin{equation}\label{usethis}
\theta_3(0,\pi e^{x})= e^{-\frac{1}{2}x}\theta_3(0,\pi e^{-x}).
\end{equation}

\begin{prop}\label{Phi_phi}
In generic ordered expression, $\Phi_*(s{\pm}\frac{1}{i\h}u{\ctt}v)$ is
a $H\!ol({\mathbb C}^2)$-valued holomorphic function on
${\rm{Re}}\,s>-\frac{1}{2}$, and 
$$
\varGamma_*(\frac{1}{2}(s{\pm}\frac{1}{i\h}u{\ctt}v)){*}
\zeta_*(s{\pm}\frac{1}{i\h}u{\ctt}v)=
e_*^{\frac{1}{2}(s{\pm}\frac{1}{i\h}u{\ctt}v)\log\pi}{*}\Phi_*(s{\pm}\frac{1}{i\h}u{\ctt}v).
$$
\end{prop}

Split the integral \eqref{phiphi} into 
$\int_{0}^{\infty}{+}\int_{-\infty}^0$,  we have 
$$
\begin{aligned}
&\int_{-\infty}^{\infty} 
\sum_{n=1}^{\infty}e^{-n^2\pi e^x} 
e_*^{\frac{1}{2}x(s{\pm}\frac{1}{i\h}u{\ctt}v)}dx=
(\int_{0}^{\infty}{+}\int_{-\infty}^{0}) 
\frac{1}{2}(\theta_3(0, \pi e^x){-}1)e_*^{\frac{1}{2}x(s{\pm}\frac{1}{i\h}u{\ctt}v)}dx\\
&=
\int_{0}^{\infty}\frac{1}{2}(\theta_3(0,\pi e^x){-}1)e_*^{\frac{1}{2}x(s{\pm}\frac{1}{i\h}u{\ctt}v)}dx
{+}
\int_{-\infty}^{0}\frac{1}{2}\theta_3(0,\pi e^{x})
e_*^{\frac{1}{2}x(s{+}\frac{1}{i\h}u{\ctt}v)}dx
{-}\frac{1}{2}\int_{-\infty}^{0}\!\!e_*^{\frac{1}{2}x(s{\pm}\frac{1}{i\h}u{\ctt}v)}dx
\end{aligned}
$$
The last term make sense for ${\rm{Re}}\,s{>}-\frac{1}{2}$.
Using \eqref{usethis} and changing variable $x$ to $-x$, the second
term becomes 
$$
\int_{-\infty}^{0}\frac{1}{2}e^{-\frac{1}{2}x}\theta_3(0,\pi e^{-x})
e_*^{\frac{1}{2}x(s{\pm}\frac{1}{i\h}u{\ctt}v)}dx{=}
\int_{0}^{\infty}\frac{1}{2}(\theta_3(0,\pi e^{x}){-}1)
e_*^{\frac{1}{2}x(1{-}s{\mp}\frac{1}{i\h}u{\ctt}v)}dx{+}
\frac{1}{2}\int_{0}^{\infty}e_*^{\frac{1}{2}x(1{-}s{\mp}\frac{1}{i\h}u{\ctt}v)}dx.
$$
The last term makes sense for ${\rm{Re}}\,s{>}\frac{1}{2}$.
Hence by assuming ${\rm{Re}}\,s{>}\frac{1}{2}$, we have 
$$
\begin{aligned}
\int_{-\infty}^{\infty} 
\sum_{n=1}^{\infty}e^{-n^2\pi e^x} 
e_*^{\frac{1}{2}x(s{\pm}\frac{1}{i\h}u{\ctt}v)}dx&{=}
\int_{0}^{\infty}\frac{1}{2}(\theta_3(0,\pi e^{x}){-}1)
(e_*^{\frac{1}{2}x(s{\pm}\frac{1}{i\h}u{\ctt}v)}{+}
e_*^{\frac{1}{2}x(1{-}s{\mp}\frac{1}{i\h}u{\ctt}v)})dx\\
&\quad{+}\int_{0}^{\infty}\frac{1}{2}e_*^{\frac{1}{2}x(1{-}s{\mp}\frac{1}{i\h}u{\ctt}v)}dx
{-}\int_{-\infty}^0\frac{1}{2}e_*^{\frac{1}{2}x(s{\pm}\frac{1}{i\h}u{\ctt}v)}.
\end{aligned}
$$
Note that the first line is entire w.r.t. $s$.  
The integrals in the last line are changed into 
$$
\int_{-\infty}^0\frac{1}{2}e_*^{\frac{1}{2}x(s{\pm}\frac{1}{i\h}u{\ctt}v)}=(s{\pm}\frac{1}{i\h}u{\ctt}v)_{*\pm}^{-1}
$$
and 
$$
\int_{0}^{\infty}\frac{1}{2}e_*^{\frac{1}{2}x(1{-}s{\mp}\frac{1}{i\h}u{\ctt}v)}dx=
\int_{-\infty}^0e_*^{x(s{-}1{\pm}\frac{1}{i\h}u{\ctt}v)}dx=
(s{-}1{\pm}\frac{1}{i\h}u{\ctt}v)_{*\pm}^{-1}={-}(1{-}(s{\pm}\frac{1}{i\h}u{\ctt}v))_{*\pm}^{-1}.
$$
Note this is symmetric by the transformation
$s{\pm}\frac{1}{i\h}u{\ctt}v$ to $1{-}(s{\pm}\frac{1}{i\h}u{\ctt}v)$.

These are analytically continued to the domain 
${\mathbb C}\setminus\{\frac{1}{2},\frac{-1}{2},\frac{-3}{2},\cdots\}$. 
Hence the formula holds on this domain, that is,  
\begin{equation}\label{symsym}
\begin{aligned}
\varGamma_*(\frac{1}{2}(s{\pm}\frac{1}{i\h}u{\ctt}v)){*}
&e_*^{-\frac{1}{2}(s{\pm}\frac{1}{i\h}u{\ctt}v)\log\pi}
\zeta_*(s{\pm}\frac{1}{i\h}u{\ctt}v)\\
=&
\int_0^{\infty}
\frac{1}{2}\big(\theta_3(0, \pi e^x)-1\big)
\big(e_*^{\frac{1}{2}x(s{\pm}\frac{1}{i\h}u{\ctt}v)}+
e_*^{\frac{1}{2}x(1{-}(s{\pm}\frac{1}{i\h}u{\ctt}v))}\big)dx\\
&\qquad {-}(1{-}(s{\pm}\frac{1}{i\h}u{\ctt}v))_{*\pm}^{-1}{-}
    (s{\pm}\frac{1}{i\h}u{\ctt}v)_{*\pm}^{-1}.
\end{aligned}
\end{equation} 

\medskip 
The similar proof of Proposition \ref{elimfactor} gives 
\begin{prop}\label{elimfactor22}
Singularities at $\{\frac{-1}{2},\frac{-3}{2},\cdots, \frac{-2n{-}1}{2}\}$ are 
eliminated in 
$$(s{+}\frac{1}{i\h}u{\ctt}v){*}\prod_{\ell=1}^n
(1{+}\frac{1}{\ell}(s{\pm}\frac{1}{i\h}u{\ctt}v)){*}
e_*^{-\frac{1}{\ell}(s{\pm}\frac{1}{i\h}u{\ctt}v)}
{*}e_*^{\frac{1}{2}(s{\pm}\frac{1}{i\h}u{\ctt}v)\log\pi}{*}\Phi_*(s{\pm}\frac{1}{i\h}u{\ctt}v).
$$ 
\end{prop}

\bigskip

Note that \eqref{symsym} is symmetric under the transformation 
$s{+}\frac{1}{i\h}u{\ctt}v \to 1-(s{+}\frac{1}{i\h}u{\ctt}v)$.  
Hence we have in generic ordered expression 
\begin{equation}\label{refsym01}
\begin{aligned}
\varGamma_*(\frac{1}{2}(s{\pm}\frac{1}{i\h}u{\ctt}v)){*}
&e_*^{-\frac{1}{2}(s{\pm}\frac{1}{i\h}u{\ctt}v)\log\pi}
{*}\zeta_*(s{\pm}\frac{1}{i\h}u{\ctt}v)\\
&=
\varGamma_*(\frac{1}{2}(1{-}(s{\pm}\frac{1}{i\h}u{\ctt}v))){*}
e_*^{-\frac{1}{2}(1{-}(s{\pm}\frac{1}{i\h}u{\ctt}v))\log\pi}
{*}\zeta_*(1{-}(s{\pm}\frac{1}{i\h}u{\ctt}v)), 
\end{aligned}
\end{equation}
on ${\mathbb C}\setminus\{\cdots,\frac{-3}{2}, \frac{-1}{2},\frac{1}{2}\}$.

\bigskip
Note that \eqref{symsym} is not symmetric under the transformation
$s\to 1{-}s$. Hence we consider 
\begin{equation}\label{symetrize}
\begin{aligned}
F_*(s){=}\varGamma_*(\frac{1}{2}(s{+}\frac{1}{i\h}u{\ctt}v)){*}
&e_*^{-\frac{1}{2}(s{+}\frac{1}{i\h}u{\ctt}v)\log\pi}
{*}\zeta_*(s{+}\frac{1}{i\h}u{\ctt}v)\\
&{+}
\varGamma_*(\frac{1}{2}(s{-}\frac{1}{i\h}u{\ctt}v)){*}
e_*^{-\frac{1}{2}(s{-}\frac{1}{i\h}u{\ctt}v)\log\pi}
{*}\zeta_*(s{-}\frac{1}{i\h}u{\ctt}v)
\end{aligned}
\end{equation}
\begin{equation}\label{symetrize2}
\begin{aligned}
G_*(s){=}\varGamma_*^{-1}(\frac{1}{2}(s{+}\frac{1}{i\h}u{\ctt}v)){*}
&e_*^{\frac{1}{2}(s{+}\frac{1}{i\h}u{\ctt}v)\log\pi}
{*}\zeta_*^{-1}(s{+}\frac{1}{i\h}u{\ctt}v)\\
&{+}
\varGamma_*^{-1}(\frac{1}{2}(s{-}\frac{1}{i\h}u{\ctt}v)){*}
e_*^{\frac{1}{2}(s{-}\frac{1}{i\h}u{\ctt}v)\log\pi}
{*}\zeta_*^{-1}(s{-}\frac{1}{i\h}u{\ctt}v)
\end{aligned}
\end{equation}
It is obvious to have $F_*(s)=F_*(1{-}s)$, and $G_*(s)=G_*(1{-}s)$. 

Consider now hybrid expressions 
$$
{:}G_{*}(s){:}_{_{E\pm}}={:}\varGamma_*^{-1}(\frac{1}{2}(s{+}\frac{1}{i\h}u{\ctt}v)){:}_{_{E(K)mat}}{+}
{:}\varGamma_*^{-1}(\frac{1}{2}(s{-}\frac{1}{i\h}u{\ctt}v)){:}_{_{\overline{E}(K)mat}}\quad
(\text{ Cf. \eqref{totresult}})
$$
$$
{:}e_*^{\frac{1}{2}(s{+}\frac{1}{i\h}u{\ctt}v)\log\pi}{:}_{_{E\pm}}=
{:}e_*^{\frac{1}{2}(s{+}\frac{1}{i\h}u{\ctt}v)\log\pi}{:}_{_{E(K)mat}}{+}
{:}e_*^{\frac{1}{2}(s{-}\frac{1}{i\h}u{\ctt}v)\log\pi}{:}_{_{\overline{E}(K)mat}}\quad 
(\text{Cf.\, Theorem\,\ref{unifiedrep}})
$$
$$
{:}\zeta_*^{-1}(s{+}\frac{1}{i\h}u{\ctt}v){:}_{_{E\pm}}=
{:}\zeta_*^{-1}(s{+}\frac{1}{i\h}u{\ctt}v){:}_{_{E(K)mat}}{+}
{:}\zeta_*^{-1}(s{-}\frac{1}{i\h}u{\ctt}v){:}_{_{\overline{E}(K)mat}},\quad
(\text{Cf. \eqref{matzetainv}}).
$$
Each of them has no singular point on ${\rm{Re}}\,s >1/2$.
We have also  
$$
G_*(s)={:}G_{*}(s){:}_{_{E\pm}}{*}{:}e_*^{\frac{1}{2}(s{+}\frac{1}{i\h}u{\ctt}v)\log\pi}{:}_{_{E\pm}}
{*}{:}\zeta_*^{-1}(s{+}\frac{1}{i\h}u{\ctt}v){:}_{_{E\pm}}
$$
and similarly
$$
G_*(1{-}s)={:}G_{*}(1{-}s){:}_{_{E\pm}}{*}
{:}e_*^{\frac{1}{2}(1{-}s{-}\frac{1}{i\h}u{\ctt}v)\log\pi}{:}_{_{E\pm}}
{*}{:}\zeta_*^{-1}(1{-}s{-}\frac{1}{i\h}u{\ctt}v){:}_{_{E\pm}}.
$$

\bigskip
By Proposition\,\ref{invzeta} both $\zeta_*(z{\pm}\frac{1}{i\h}u{\ctt}v)$
have genuine inverses for ${\rm{Re}}\,z>\frac{1}{2}$. 

We now take the hybrid matrix expression of the identity 
$G_*(s)=G_*(1{-}s)$.

Note also  that if a function $f(X)$ has no singular point on 
the domain ${\rm{Re}}X>\frac{1}{2}$, if $f(X)$ has the 
reflection symmetric property such as 
$f(X)=f(1-X)$, then the singular points of $f(X)$ 
possibly allows only on the line ${\rm{Re}}X=\frac{1}{2}$.

The next result might suggest useful information about Riemann conjecture.
\begin{thm}\label{Riemann}
Singular point of ${:}\zeta_*(s{+}\frac{1}{i\h}u{\ctt}v)^{-1}{:}_{_{E\pm}}$ on the strip 
$0{<}{\rm{Re}}\,s{<}1$ is possibly only on ${\rm{Re}}\,s{=}\frac{1}{2}$. 
\end{thm}


\begin{thebibliography}{OM}
 
\bibitem{AAR}{G. Andrews, R. Askey, R. Roy},\,\,
\newblock{\sc Special functions},
\newblock{Encyclopedia Math, Appl. 71, Cambridge, 2000.}


\bibitem{AW}{G. S. Agawal, E. Wolf},\,\,
\newblock{\it Calculus for functions of noncommuting 
operators and general phase-space method of functions},
\newblock{Physcal Review D, vol.2, no.10, 1970, 2161-2186.}


\bibitem{BF}{F.Bayen,\,M,Flato,\,C.Fronsdal,\,A.Lichnerowicz,\,D.Sternheimer,\,}
{\it Deformation theory and quantization I, II}, Ann. Phys. 111, (1977),
61-151.  


\bibitem{GS}{I.M.Gel'fand,\,G.E.Shilov,\,\,}
{Generalized Functions, 2}, Acad. Press, 1968. 


\bibitem{gus}{V. Guillemin, S. Sternberg},\,\, 
{\sc Geometric Asymptotics}. A.M.S. Mathematical surveys, 14, 1977. 


\bibitem{Hi}{N. Hitchin},
\newblock{\it Lectures on special Lagrangian submanifolds}, 
\newblock{arXiv:math.DG/9907034vl 6Jul, 1999.}


 
\bibitem{Mm}{M.Morimoto,\,\,}{\sc An introduction to Sato's hyperfunctions},
 AMS Trans. Mono.129, 1993.  

\bibitem{om3}{H. Omori},\,\,{\sc Infinite dimensional Lie groups}, AMS
  Translation Monograph 158, 1997.   

\bibitem{om6}{H. Omori},\,\,{\it Toward geometric quantum theory}, in 
From Geometry to Quantum Mechanics. Prog. Math. 252,
Birkh{\"a}user,2007, 213-251.    


{\bibitem{ommy}H. Omori, Y. Maeda, N. Miyazaki, A. Yoshioka}, 
\newblock{\it Deformation quantization of Fr\'echet-Poisson algebras, 
--{Convergence of the Moyal product--}},  
in  Conf\'erence Mosh\'e Flato 1999, Quantizations, Deformations, 
and Symmetries, Vol II,  Math. Phys. Studies 22, 
Kluwer Academic Press, 2000, 233-246.


{\bibitem{ommy}H. Omori, Y. Maeda, N. Miyazaki, A. Yoshioka}, 
\newblock{\it Strange phenomena related to ordering problems in 
quantizations},  Jour. Lie Theorey, Vol 13, no.2, (2003) 491-510.



\bibitem{OMMY6}
{H.Omori, Y.Maeda, N.Miyazaki and A.Yoshioka :}
\newblock{\em Star exponential functions as two-valued elements}, 
  \newblock{in The breadth of symplectic and Poisson geometry}, 
  \newblock{Progress in Math. 232, Birkh{\"auser},2004}, 483-492. 



\bibitem{OMMY7}
{H.Omori, Y.Maeda, N.Miyazaki and A.Yoshioka :}
\newblock{\em Strange phenomena related to ordereing problems in quantizations},  
  \newblock{Jour. Lie Theory, Vol 13, No 2, (2003)}, 481-510. 


\bibitem{OMMY3}
{H.Omori, Y.Maeda, N.Miyazaki and A.Yoshioka :}
\newblock{\it Deformation of expressions for elements of algebras  (I)}, 
 -(Jacobi's theta functions and $*$-exponential functions)-
{arXiv:1104.2109} 
 

\bibitem{OMMY4}
{H.Omori, Y.Maeda, N.Miyazaki and A.Yoshioka :}
\newblock{\it Deformation of expressions for elements of algebras  (II)}, 
-(Weyl algebra of $2m$-generators)-
{arXiv:1105.1218} 


\bibitem{OMMY5}
{H.Omori, Y.Maeda, N.Miyazaki and A.Yoshioka :}
\newblock{\it Deformation of expressions for elements of algebras  (III)}, 
-Generic product formula for ${*}$-exponentials of quadratic forms-
{arXiv:1107.2474}


\bibitem{ommy6}
{H.Omori, Y.Maeda, N.Miyazaki and A.Yoshioka :}
\newblock{\it Deformation of expressions for elements of algebras  (IV)}
-Matrix elements and related integrals-
{arXiv:1109.0082} 



 

%
%
%


 
%
\end{thebibliography}
\end{document}